\documentclass[12pt,preprint]{aastex}

\begin{document}
\parindent=1.0cm

\title{Coaxing the Eclipsing Binary V367 Cygni Out of its Shell}

\author{T.J. Davidge\altaffilmark{1,2}}

\affil{Dominion Astrophysical Observatory,
\\Herzberg Astronomy \& Astrophysics Research Center,
\\National Research Council of Canada, 5071 West Saanich Road,
\\Victoria, BC Canada V9E 2E7\\tim.davidge@nrc.ca; tdavidge1450@gmail.com}

\altaffiltext{1}{Based on observations obtained at the Dominion Astrophysical 
Observatory, Herzberg Astronomy and Astrophysics Research Centre, National 
Research Council of Canada.}

\altaffiltext{2}{This research has made use of the NASA/IPAC Infrared Science 
Archive, which is funded by the National Aeronautics and Space Administration 
and operated by the California Institute of Technology.}

\begin{abstract}

	Spectra that cover $0.63 - 0.69\mu$m with a spectral resolution 
$\sim 17000$ are presented of the W Serpentis system V367 
Cygni. Absorption lines of FeII and SiII that form in a 
circumsystem shell are prominent features, and the depths of these 
are stable with time, suggesting that the shell is 
smoothly distributed and well-mixed. Further evidence of uniformity 
comes from modest radial velocity variations measured in the 
deepest parts of the shell lines. It is suggested that motions previously 
attributed to rotation of the shell are instead artifacts of 
contamination from the donor star spectrum. A donor star spectrum 
is extracted that is consistent with that of an early to mid-A giant. 
The depths of metallic lines in the donor spectrum vary with orbital 
phase, suggesting that spot activity covers 
a large fraction of the surface of that star. 
A spectrum of the accretion disk that surrounds the second 
star is also extracted, and similarities are noted with the emission 
spectra of Herbig Ae/Be stars. In addition to variations with orbital phase, 
H$\alpha$ changes with time over timescales 
of no more than two orbits. A tentative detection of HeI 6678 
emission is made near primary minimum, but not at 
other phases. Finally, projected emission from hot dust 
in and around V367 Cyg is more-or-less symmetric and extends over 
28 arcsec, or 0.09 pc at the distance of the system -- V367 Cyg is thus 
expelling matter into a large volume of the surrounding space.

\end{abstract}

\section{INTRODUCTION}

	Many stars in the solar neighborhood are in 
binary systems (e.g. Abt 1983). If the stars are close enough together 
then mass will be transfered between components. Initially, material from the 
more massive star is transfered to the lower mass star as evolution 
causes the former to expand and fill its Roche lobe. An additional 
exchange of mass may occur when the star that accretes this material evolves. 

	The transfer of mass and angular momentum in such close 
binary systems (CBSs) causes the evolution of both components 
to depart from that of single stars, and this has potential 
implications for a range of astrophysical topics. 
Departures from single star evolution are expected to be 
common among massive stars, as the binary frequency among these objects 
is high (Sana et al. 2012). Such interactions in massive CBSs can skew 
population statistics in studies of the stellar contents of other galaxies 
(De Mink et al. 2014), and it has been suggested that some 
of the brightest objects in star-forming galaxies, such as luminous 
blue variables and supergiant B[e] stars, may be CBSs 
(e.g. Smith \& Tombleson 2015; Clark et al. 2013). 
Mass transfer can also affect the properties of the compact objects 
that are the end products of massive star evolution (e.g. Ertl et al. 2020).

	As with massive stars, a significant population of intermediate 
mass stars are likely in CBSs. Semi-detached (SD) systems, in which 
one star fills its Roche lobe, are common among eclipsing 
binaries that contain intermediate mass stars 
(e.g. Malkov 2020). Given that eclipsing systems sample a restricted range 
of orbital orientations then the number of as yet undetected intermediate 
mass SD systems will be even higher than in the 
Malkov (2020) census. One implication is that 
binary white dwarf (WD) systems, which are one possible end point 
of CBS evolution involving intermediate mass stars, may not be rare. 

	Systems that are in the early stages of interaction are of particular 
interest as the component stars are closer to their original states than in 
more evolved systems. Such systems thus require fewer 
assumptions about the nature of the progenitor stars when making 
comparisons with models. That being said, the rate of mass transfer is highest 
at these times, as evolution forces the donor 
star to expand into its Roche lobe and the separation between 
components decreases prior to the reversal of the mass ratio.
The rapid early evolution of these systems can 
complicate efforts to understand the system properties, as an accretion disk 
around the accreting star may block light from that object, while a 
circumsystem shell may complicate efforts to detect the spectra of 
both components. Models of CBSs discussed by Nelson \& Eggleton (2001) predict 
a large initial spike in mass transfer rates over a dynamical timescale, 
followed by a mass transfer rate that is $\sim 2$ orders of magnitude 
lower than the peak value. 

	W Serpentis (W Ser) is an enigmatic CBS that is the archetype of a 
group first discussed by Plavec \& Koch (1978). 
Aside from binarity, a defining characteristic of this group is a UV 
spectrum that is indicative of excitation levels that can not be attributed to 
the component stars. There are also photometric 
and spectroscopic properties that are suggestive of large scale mass 
transfer. $\beta$ Lyrae is likely the nearest W Ser 
system, and interferometric observations discussed by Zhao et al. (2008) reveal 
the distorted nature of its components. 

	W Ser systems are thought to be in the early phases of mass transfer. 
The stream of matter from the donor star forms an accretion 
disk. A hot spot may then form where the stream impacts 
the disk, and this spot can affect the observational properties of the 
system. Table 1 of Van Rensbergen et al. (2011) lists 
spot temperatures in selected W Ser systems, and spots with 
temperatures in excess of 20000K are common. Hot spots thus provide a 
plausible source of the high energies needed to power UV emission. 
Variations in the spot brightness and location may also affect the 
photometric properties of W Ser systems at visible wavelengths. 

	Only a modest number of W Ser systems have been identified to date, 
and the majority of these are located within $\sim 1$ kpc (e.g. Elias 1990). 
The small number of confirmed systems is in part a selection effect, as 
there is a bias towards eclipsing systems (i.e. those with orbital inclinations 
in excess of $\sim 65^o$). There are undoubtedly 
non-eclipsing W Ser systems located within 1 kpc that remain to be 
identified or that have an uncertain classification.

	In the current paper we examine the spectroscopic 
properties of V367 Cyg, which is one of the five systems 
discussed by Plavec \& Koch (1978). The parallax of V367 
Cyg listed in the GAIA Data Release 2 (Brown et al. 2018) is 1.05 
marcsec, indicating a distance of 670 parsecs. \footnote[1]{This distance 
has not been corrected for possible systematic effects in the parallax 
measurements, such as zeropoint offsets (e.g. Lindgren et al. 2018).}
Models examined by Van Rensbergen et al. (2011) suggest that 
V367 Cyg is likely undergoing Case B mass transfer given its orbital 
period of 18.6 days. UV emission line strengths suggest that processed 
material has not yet been accessed in the donor star (Plavec et al. 1984). 

	The broad-band spectral energy distribution (SED) of V367 Cyg 
provides basic limits on the effective temperatures of the 
system components. Based on a single low resolution spectrum 
although this classification was undoubtedly influenced by shell lines, and so 
does not reflect the actual spectral types of the component stars. 
The UV continuum is consistent with sources that have 
effective temperatures of 8000 and 10000K (Hack et al. 1984), 
while the SED between 1 and $2.5\mu$m is indicative of components that
have effective temperatures of 9000 and 11000 K (Taranova \& Shenavrin 2005). 
The temperatures found from the UV and IR data thus suggest that the dominant 
components have SEDs that are consistent with late B/early A to 
mid-A spectral types. Still, the light from the accreting star may be 
masked by an accretion disk. Moreover, at longer wavelengths the 
SED contains excess infrared (IR) emission (Taranova 1997, Taranova 
\& Shenavrin 2005), indicating a non-stellar contribution to the 
SED. The IR properties are similar to those of 
$\beta$ Lyrae (Taranova \& Shenavrin 2005), although there are differences 
at radio wavelengths (Elias 1990). 

	It has been known for some time that the dominant 
absorption lines in the spectrum of V367 Cyg at visible 
wavelengths originate in a circumsystem shell (e.g. Abt 1954; Heiser 1961 
and references therein). The shell lines at wavelengths $< 5000$\AA\ 
tend to have similar kinematic behaviours, although there are 
exceptions (Aydin et al. 1978; Hack et al. 1984) that have been attributed 
to ejection events (Hack et al. 1984). The profiles of the shell line vary with 
orbital phase, and this has been linked to possible non-uniform shell 
structure (Abt 1954, Heiser 1961, Aydin et al. 1978). Variations related to 
ionization are also present that are suggestive 
of stratification (Aydin et al. 1978). Hack et al. (1984) 
estimate a shell mass of 10$^{-5}$ to 10$^{-6}$ M$_{\odot}$.

	Absorption lines in the spectrum of the donor star have been 
detected in previous studies, although veiling by the shell, 
coupled with line broadening due to the donor's rotation, complicate 
efforts to detect these lines. Even though shell 
lines may mask the donor spectrum, lines from the donor can still 
be untangled from shell lines given that the orbital velocity of the donor 
differs markedly from that of the shell (Section 5). While high order Balmer 
lines and Ca H and K show evidence of a contribution from the donor star 
(Heiser 1961, Aydin et al. 1978), past attempts to examine the properties 
of the donor star have largely focused on MgII 4481.

	Despite extensive searches, no spectroscopic signatures 
of the accreting star have been found (Schneider et al. 1993). 
This is perhaps not surprising, as the accreting star is intrinsically 
fainter than the more evolved donor, and an accretion disk may mask 
much of its light. V367 Cyg is thus a single line spectroscopic 
binary, and so the masses of the component stars are uncertain. This 
uncertainty is reflected in the published mass estimates, 
which are summarized in Table 1. The uncertainties in this table are from 
the source papers, when available. There is considerable 
dispersion among the entries, highlighting the uncertain nature of V367 Cyg and 
its eventual end state. 

\begin{table*}[ht]
\begin{center}
\begin{tabular}{cccl}
\tableline\tableline
Donor & Accretor & Source & Method \tablenotemark{a}\\
(M$_{\odot}$) & (M$_{\odot}$) & & \\
\tableline
6.7 & 6.7 & Aydin et al. (1978) & RVs without disk; mass ratio fixed \\
$19 \pm 4$ & $12 \pm 3$ & Li \& Leung (1987) & LC fitting without disk; no RVs \\
4.4 & 2.5 & Pavlovski et al. (1992) & LC+RVs; with disk \\
22 & 11 & Tarasov \& Berdyugin (1998) & LC+RV of accretor based on HeI 6678 \\
$21.7 \pm 2.8$ & $11.0 \pm 0.8$ & Zola \& Ogloza (2001) & LC without disk; l$_3 = 0$ \\
$3.3 \pm 0.9$ & $4.0 \pm 0.5$ & Zola \& Ogloza (2001) & LC with disk; l$_3 = 0$ \\
\tableline
\end{tabular}
\end{center}
\caption{Mass estimates}
\tablenotetext{a}{RV = radial velocity; LC = light curve; l$_3$ = third light}
\end{table*}

	While the detection of spectroscopic signatures from the accreting star 
is problematic, there are other potential tracers of its orbital motion.
For example, given a possible association with the hot spot, UV 
emission lines might probe the dynamical properties of the accretor and 
the surrounding disk, including the escape velocity (Deschamps et al. 2015). 
Tarasov \& Berdyugin (1998) discuss the possible detection 
of HeI 6678 emission, pointing out that it defines a 
velocity curve that is out of phase with that of the 
donor. They argue that the line originates in the accretion disk that 
surrounds the accreting star, and use it to trace the 
orbital motion of the latter. The detection of HeI 6678 has been challenged 
by Zola \& Ogloza (2001), who note that the detected feature is in a 
complex mix of shell lines, making its identification uncertain.

	Additional uncertainties in the nature of the stars in V367 
Cyg result from the complicated nature of the light curve. 
The light curves of W Ser systems tend to have a unequal maxima. This 
is not a rare phenomenon among CBSs (e.g. O'Connell 1951, Milone 1968, Davidge 
\& Milone 1984), and there are indications that there is more than one 
cause (e.g. Davidge \& Milone 1984). The differences between the 
out-of-eclipse maxima in W Ser systems tend to vary over time scales of many 
orbital cycles, and this is also the case in the V367 Cyg light curve. Such 
variations highlight the non-uniform distribution of light-emitting regions 
in the system, likely due in part to the accretion hot spot and possible 
activity on the donor star. Such photometric irregularities complicate 
light curve analysis.

	The accretion disk is a significant source of light in V367 Cyg that 
must be considered in light curve models. Perhaps the most direct 
approach is to consider the accreting star and its accompanying disk as 
a single star-like body (i.e. assume that the disk hugs the surface of the 
accreting star). Solely in terms of light output this is not a bad 
assumption, as the accreting star may be completely shielded by the disk. 
However, the shape of a rapidly rotating disk will almost 
certainly differ from that of a star, compromising the physical interpretation 
of parameters estimated from the light curve. Li \& Leung (1987) and Zola 
\& Ogloza (2001) examine such models and find that the donor star and the other 
body are in, or are close to, contact with their Roche surfaces. 
An inspection of Table 1 indicates that these light curve solutions also 
tend to predict component masses that are at the upper 
end of published values. Li \& Leung (1987) further note that 
the light curve solution changes with epoch, and 
attribute this to variations in the degree of contact. However, 
it is more likely that changes in the light curve over 
timescales of a few years are due to variations in the 
placement, intensity, and/or physical extent of a hot spot, rather than 
physical changes in the system geometry. 

	Pavlovski et al. (1992) and Zola \& Ogloza (2001) 
examine model light curves for V367 Cyg that include an accretion 
disk with a geometry that does not follow an equipotential 
surface for a static, non-rotating object. The 
models have difficulty matching the observations 
near secondary minimum, hinting at a light distribution in the disk that is 
more complicated than that adopted for the model. Wilson (1981; 2018) 
discusses models for systems that have large-scale accretion/decretion disks. 
Wilson (2018) generates a model light curve for a system with parameters 
like those expected for $\beta$ Lyrae, and demonstrates that a large 
fraction of the variation in the light curve can be attributed to the disk. In 
the current paper it is shown that emission lines at visible wavelengths are 
strongest and also vary in strength with time near secondary minimum, 
which is the phase where the disk is eclipsed.

	Zola \& Ogloza (2001) conclude that the disk 
contributes $\sim 21\%$ of the system light in $I$, and 
estimate a mass transfer rate of $5 - 7 \sim 10^{-5}$ M$_{\odot}$/year. 
This rate of mass transfer suggests that the mass ratio of the component stars 
will reverse on time scales of $< 10^5$ years if the lower mass estimates 
in Table 1 are correct. This timescale for mass reversal 
is roughly an order of magnitude faster than predicted for the systems 
modelled by Nelson \& Eggleton (2001). 

	Another potential complication when modelling the V367 Cyg light curve 
is that there are other stars that contribute to the light. There is a 
companion that is $\sim 5$ magnitude fainter and 2.5 arcsec distant, and there 
are hints of another companion that is only $\sim 0.15$ arcsec 
($\sim 300$ AU) distant (e.g. Heiser 1961 and references 
therein, McAlister \& Hattkopf 1988). Both of these objects fall within the 
plausible separation regime for physical companions (e.g. 
Simon 1997). Nelson \& Eggleton (2001) suggest that 
companion stars and the resulting impact on 
angular momentum transfer are a possible explanation for the failure of their 
models to reproduce the properties of some Algol systems.

	Light curve solutions have the potential to provide 
observational limits on the brightness of unresolved components 
via the inclusion of third light in the models (e.g. Davidge \& Forbes 
1988). This being said, Li \& Leung (1987) find no evidence for third 
light in V367 Cyg light curves, while Zola \& Ogloza (2001) find that models 
that include a disk provide only loose limits on third 
light. A close companion may explain at least some of the longer term 
photometric and spectroscopic variations that have been detected. 

	The examination of lines over a broad wavelength range 
from V367 Cyg is of interest to further explore the system properties. However, 
most spectroscopic studies of V367 Cyg have been at wavelengths 
shortward of $0.5\mu$m. In the present paper we discuss spectra of V367 
Cyg that were recorded over seven months in 2021. The 
spectra span a $0.6\mu$m wide wavelength interval centered on H$\alpha$ 
with a resolution $\frac{\lambda}{\Delta \lambda} \sim 17000$. 
With the exception of the spectra discussed by Tarasov \& Berdyugin (1998) 
that cover only a few phase intervals and have narrow 
wavelength coverage centered on specific features, 
these wavelengths have not yet been examined in the spectrum of V367 Cyg. 

\section{OBSERVATIONS AND REDUCTION}

	The spectra were recorded with the 
1.2 meter telescope and McKellar coud\'e spectrograph 
(Monin et al. 2014) at the Dominion Astrophysical Observatory (DAO). 
The detector was the $2048 \times 4088$ pixel SITe--4 CCD. This 
detector has $15\mu$m square pixels and was binned 
perpendicular to the dispersion axis during read-out, with the 
level of binning set by other programs observed 
on a given night. The spectrograph was configured with 
the IS32R image slicer, the 32 inch camera, and the 
1200H grating. The central wavelength was set near 6600\AA\ . This 
configuration produces a dispersion of 0.15 \AA / pixel, with a 
resolution $\lambda$/$\Delta\lambda \sim 17000$. 

	The spectra were recorded with the telescope operating 
in robotic mode. Cloud cover was monitored with a wide-field camera, 
and the archived images indicate that some of the 
spectra were recorded during non-photometric conditions. The cloud 
cover on the non-photometric nights was such that the S/N ratio 
was not seriously compromised. 

	Three 300 sec exposures of V367 Cyg were recorded on each night, 
and the resulting 900 sec total exposure time produced a 
S/N ratio $\geq 100$/pixel in the continuum. Heliocentric Julian Dates 
(HJD) on which spectra were recorded and the orbital 
phase at the time of observation are listed in Table 
2. The spectra cover a full range of orbital phases, with the 
densest coverage near secondary minimum. 

	Phases were calculated using the emphemeris given by Lloyd (2018). The 
O--C diagram in Figure 5 of Lloyd (2018) shows a dispersion 
of one to two tenths of a day ($\sim 0.01$ in orbital 
phase). This scatter may be due to transient events 
in the light curve, coupled with the comparatively long period of the system, 
which is such that eclipses occur over many hours, thereby complicating efforts 
to measure times of minimum on some nights. 

\begin{table*}
\begin{center}
\begin{tabular}{cccc}
\tableline\tableline
Date & Orbital Phase \tablenotemark{a} & Date & Orbital Phase \tablenotemark{a} \\
(HJD - 2459000) &  & (HJD - 2459000) & \\
\hline
289.4589 & 0.45 & 417.2137 & 0.36 \\
309.4723 & 0.56 & 419.4698 & 0.48 \\
310.4730 & 0.62 & 420.4528 & 0.53 \\
331.4905 & 0.75 & 421.2097 & 0.57 \\
356.4316 & 0.09 & 422.2553 & 0.63 \\
363.4322 & 0.46 & 437.1947 & 0.43 \\
366.4206 & 0.62 & 438.3811 & 0.49 \\
382.2732 & 0.48 & 439.2610 & 0.54 \\
383.2663 & 0.53 & 440.2825 & 0.60 \\
384.4197 & 0.59 & 441.2248 & 0.65 \\
385.4529 & 0.65 & 444.2518 & 0.81 \\
386.2690 & 0.69 & 445.1818 & 0.86 \\
387.2709 & 0.74 & 446.4072 & 0.92 \\
388.4228 & 0.81 & 468.1675 & 0.09 \\
411.2179 & 0.03 & 470.2428 & 0.21 \\
413.2834 & 0.14 & 471.1476 & 0.26 \\
415.4502 & 0.26 & 481.1289 & 0.79 \\
416.4374 & 0.31 & & \\
\tableline
\end{tabular}
\end{center}
\caption{Log of Observations}
\tablenotetext{a}{Based on the emphemeris calculated by Lloyd (2018).}
\end{table*}

	A series of calibration frames were also recorded, and these were used 
to remove instrumental signatures from the spectra. The calibration suite 
includes biases, dispersed light from a continuum source to monitor 
flat-field variations, and spectra of a Th-Ar arc for wavelength 
calibration. The spectrophotometric standard star HR8634 ($\zeta$ Peg; 
Hamuy et al. 1992) was observed on some nights.

	Processing was done with a standard pipeline for 
single slit spectra. The reduction sequence included 
bias subtraction, flat-fielding, wavelength 
calibration, and a correction to helio-centric velocities. The final step 
was the removal of variations in the continuum level to produce spectra 
with a pseudo-continuum normalized to unity.

\section{THE RED SPECTRUM OF V367 CYG: AN OVERVIEW}

	The spectra were sorted according to orbital phase, and then binned 
in ten equally wide phase intervals. A mean spectrum was constructed 
for each bin to allow the properties of the spectra throughout a 'typical' 
orbital cycle in 2021 to be examined. These mean spectra 
are the subject of more detailed examination later in the paper. 
In this section, features are highlighted in the mean spectra 
at three phases to serve as an introduction to the more detailed analysis. 
This is done as the spectra cover wavelengths that have been little studied 
in V367 Cyg. Phase 0.5 (secondary minimum) is of particular interest 
given difficulties matching the light curve there (Pavlovski et al. 
1992; Zola \& Ogloza 2001).

	The mean spectra at phases 0.3, 0.5 and 0.7 in 
three wavelength intervals are compared in Figures 1 -- 3. 
The deep absorption lines in Figures 1 and 3 originate in the 
circumsystem shell. Variations in the shell lines that are related to 
orbital phase have been documented in past studies 
at shorter wavelengths (e.g. Heiser 1961; Aydin et al. 1978), and 
phase-related variations in line morphology are evident among 
the shell lines in our spectra. The SiII 6347 doublet is one of the strongest 
shell lines at these wavelengths. This is a resonance transition, 
and its behaviour with phase is not remarkable. Although 
not obvious in Figures 1 and 3, it is argued in Sections 4 and 5 that 
absorption lines from the donor star overlap with the shell lines, and that 
these contribute to the variations in the shell line profiles with orbital 
phase. 

\begin{figure}
\figurenum{1}
\plotone{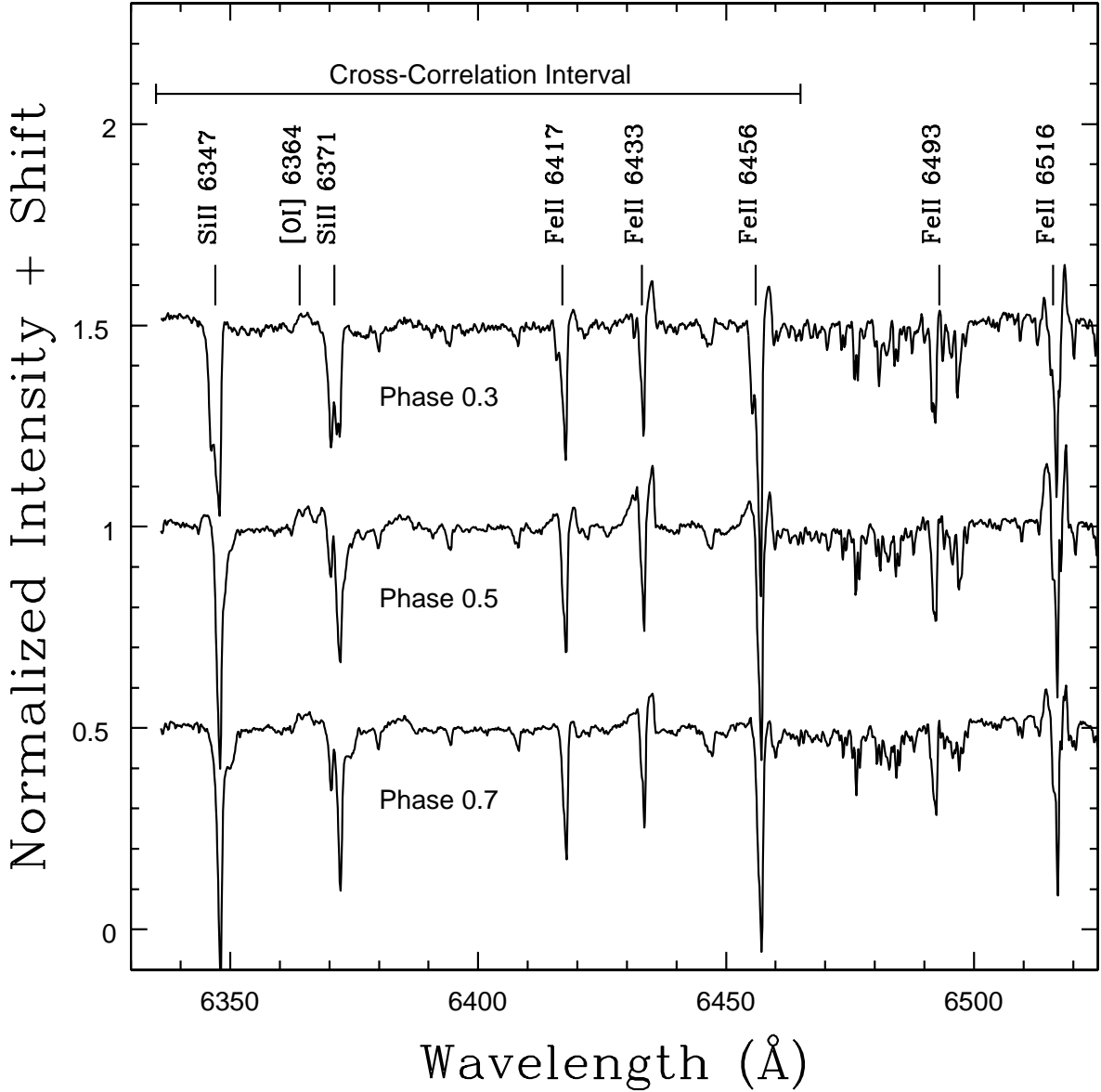}
\caption{Binned spectra of V367 Cyg at three orbital phases in the wavelength 
interval 6330 -- 6525\AA\ . The spectra have been normalized to the continuum, 
and the results have been shifted vertically for display purposes. The main 
absorption components of the SiII and FeII lines originate in the circumsystem 
shell, and it is shown in Section 5 that there are transient 
absorption features in the shell lines that originate in 
the spectrum of the donor star. The emission features in the shoulders of 
the shell lines change with phase in a manner that is not consistent 
with a shell origin, and are strongest at phase 0.5, when the accreting 
star is eclipsed. [OI] 6364 is a broad, low amplitude emission feature, with a 
shape that varies with orbital phase. The wavelength interval used for the 
cross-correlation velocity measurements in Section 4 is indicated at the 
top of the figure.}
\end{figure}

\begin{figure}
\figurenum{2}
\plotone{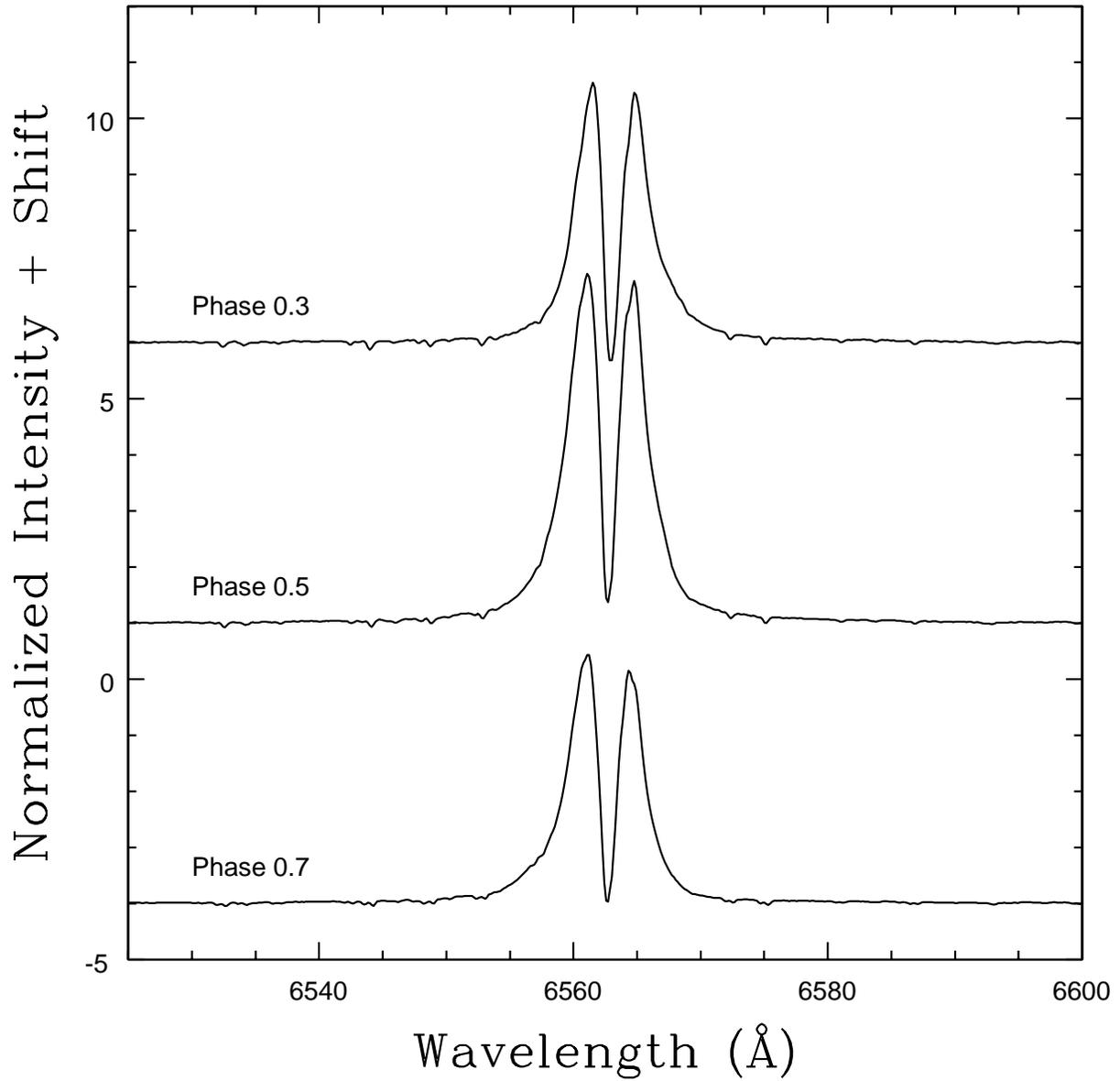}
\caption{Same as Figure 1, but covering the wavelength interval 6525 -- 
6600\AA\ . The dominant feature is H$\alpha$, and the 
strengths of the emission and absorption components change 
with orbital phase. These components are most pronounced near 
primary minimum (not shown here) and secondary minimum.}
\end{figure}

\begin{figure}
\figurenum{3}
\plotone{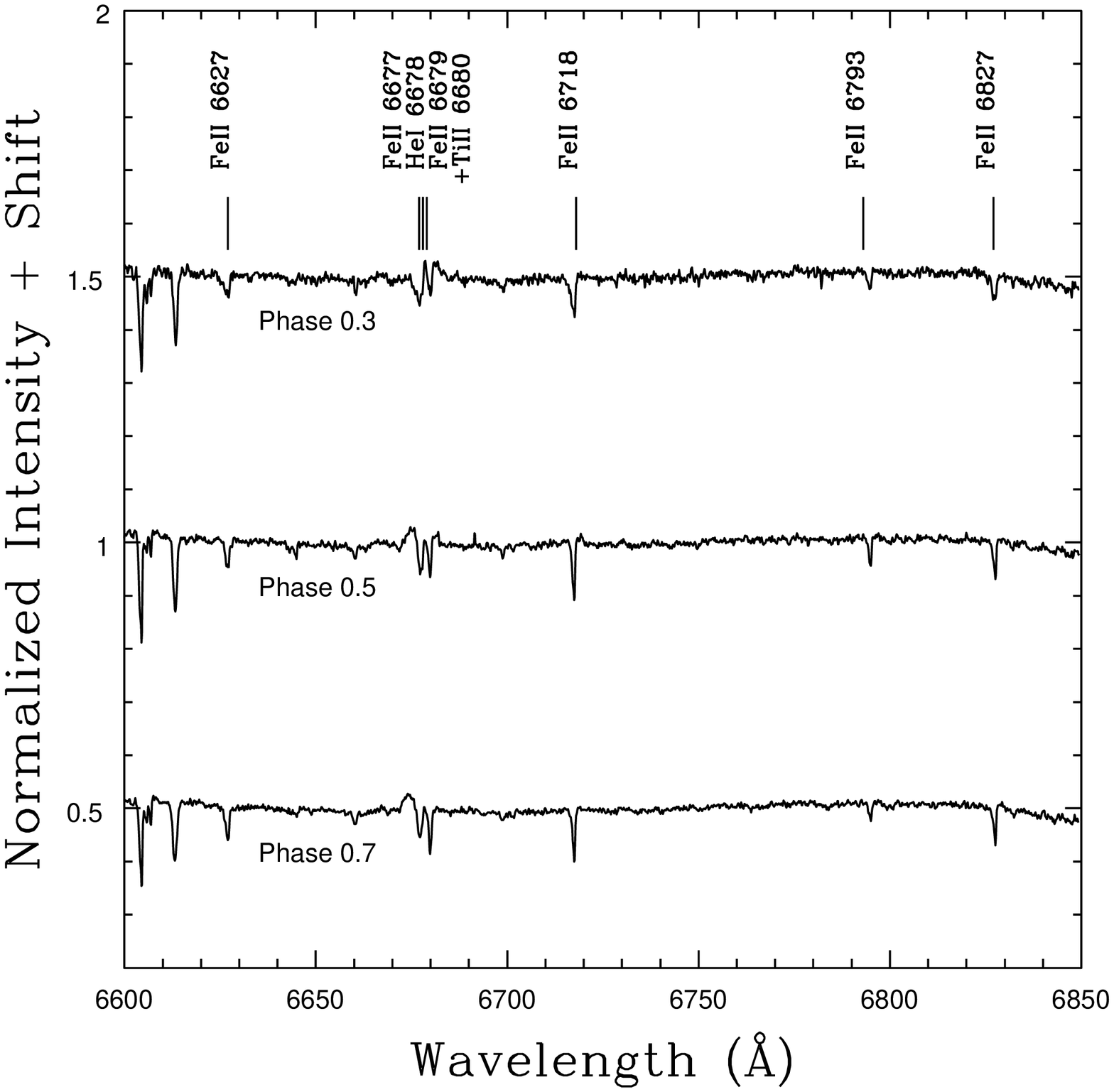}
\caption{Same as Figure 1, but covering the wavelength interval 6600 -- 
6850\AA\ . The shell lines at these wavelengths are weaker than those in 
Figure 1, but have similar transient features. HeI 6678 is located in a region 
that is dominated by shell absorption lines.}
\end{figure}

	Broad emission features are evident in the shoulders of the FeII 
and SiII shell lines, and these vary with orbital phase. The emission lines 
associated with different FeII lines vary in sync 
and move in the opposite direction to that 
expected for an object associated with the donor star. FeII emission is 
most pronounced in the blueward shoulders of the shell lines 
near phase 0.5, when the accreting star and its disk is eclipsed. In Section 6 
it is argued that these emission lines originate in the accretion disk.

	There are other, less obvious, features in Figures 1 - 3 that are not 
related to the shell. The broad, low amplitude emission feature to the left 
of the SiII 6371 doublet in Figure 1 is [OI] 6364. The broad nature 
of this feature suggests that the emission forms 
in an environment that spans a range of velocities. 
In Section 6 it is shown that there is sub-structure that is 
related to orbital phase, and it is argued that there are two sources 
of [OI] 6364 emission. HeI 6678 in Figure 3 falls 
close to FeII and TiII shell lines that blend together at this wavelength 
resolution, and an obvious signature of HeI is not seen. 

	H$\alpha$ is the most conspicuous feature in the wavelength interval 
examined in this study, and absorption and emission components are 
evident in Figure 2. The character of H$\alpha$ changes with orbital phase 
in a manner that differs from that seen in the metallic lines discussed 
above. Phase-related variations in the characteristics of H$\alpha$ 
have been discussed by Tarasov \& Berdyugin (1998), although with restricted 
phase coverage. The absorption and emission components in Figure 2 
are most pronounced near phase 0.5; while not shown here, H$\alpha$ at phase 
0.0 is similar in shape and strength to what is seen near phase 0.5 (Section 
6). While higher order Balmer lines show characteristics that are suggestive 
of an origin in the donor star (e.g. Heiser 1961; Aydin et al. 1978), 
it is demonstrated in Section 6 that the donor star makes a negligible 
contribution to H$\alpha$ absorption.

\section{THE SHELL SPECTRUM}

\subsection{The Shapes and Radial Velocities of Shell Lines}

	The morphologies of selected deep shell lines are examined 
in Figure 4, where phase-binned mean spectra of 
SiII 6347, SiII 6371, and FeII 6432 are shown. 
Phase-related sub-structuring is clearly evident in the 
line profiles, and this is most obvious in the SiII transitions. The 
most noticeable differences with orbital phase occur blueward of the 
line centers.

	The SiII 6347 doublet is a resonance transition, and 
Tarasov \& Bergyugin (1998) suggest that sub-structure in the 
SiII 6347 profile originates in a stream between the two stars. 
However, the variations in SiII 6347 with phase are mirrored 
in other Si and Fe lines - they are not restricted to SiII 6347.
In Section 5 it is shown that a large part of the variations in the 
morphology of the shell lines are linked to the donor 
star and its radial velocity variation with orbital phase.

\begin{figure}
\figurenum{4}
\plotone{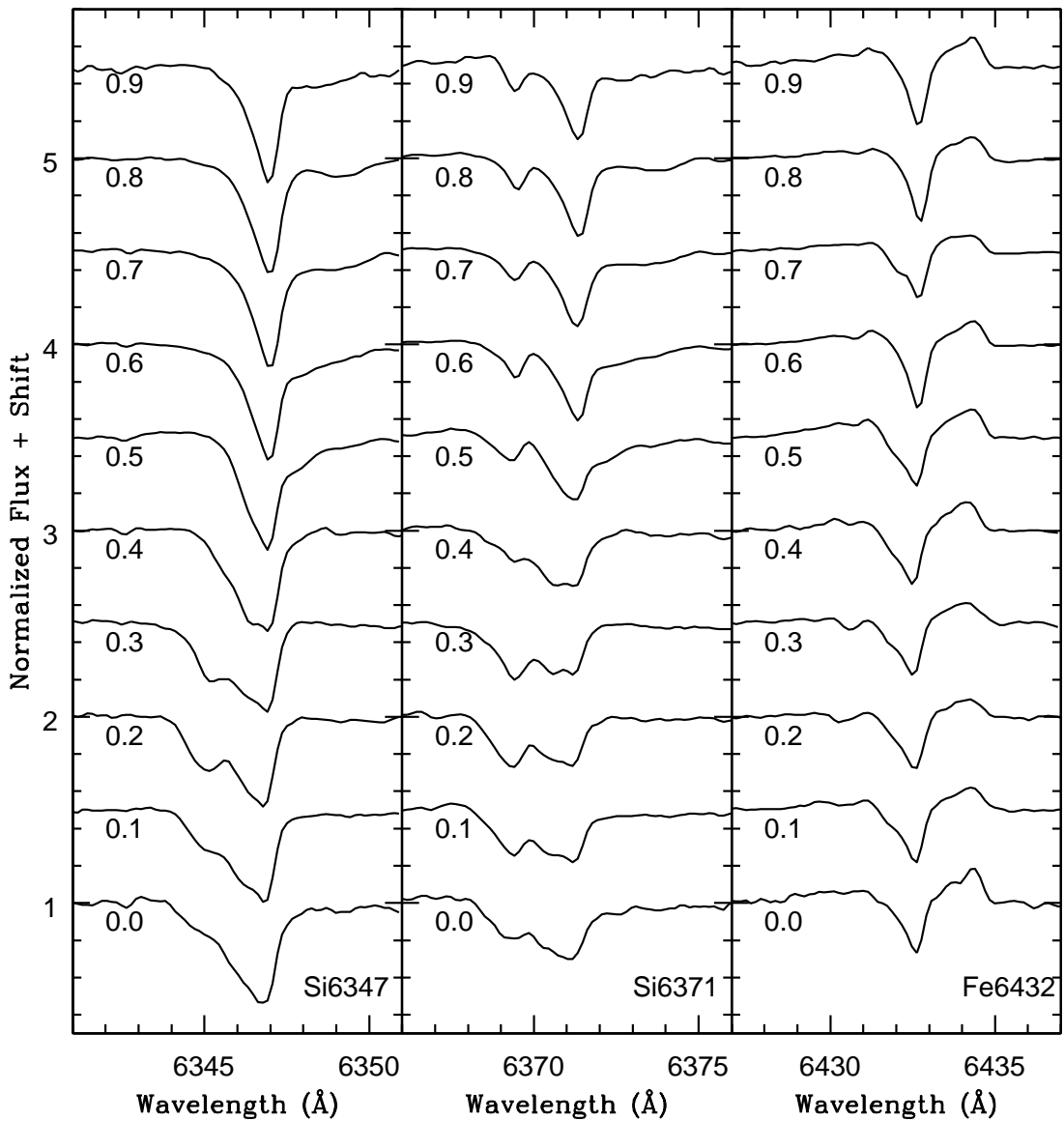}
\caption{Mean phase-binned spectra centered on SiII 6347, SiII 6371, and 
FeII 6432. Orbital phases are listed near the left edge 
of each panel. Sub-structures propogate with phase through all three 
lines, and these are most evident at wavelengths blueward of the 
line centers. It is argued in this paper that much of this sub-structure 
is associated with the donor star. The emission associated 
with FeII 6432 is more pronounced than for the SiII lines, and it is 
argued that this emission originates in an accretion disk.}
\end{figure}

	There is prominent emission in the redward shoulder of the 
FeII 6432 line, with weaker emission near the blue shoulder. 
The emission associated with FeII 6432 varies 
in wavelength with orbital phase, in the opposite 
sense to that expected for an object moving in sync with the donor star.
The emission associated with this FeII line is broad and 
may not be centered on the absorption component. Moreover, 
it is not clear if the emission originates from a 
single source (i.e. there is no structure in the emission line), or if there 
are sub-components, such as is seen in [OI] 6364 (Section 6). Emission is 
also seen in the shoulders of the two SiII lines, although the relative 
strength of the SiII emission with respect to the depth of the absorption 
feature is lower than for FeII 6432.

	Previous studies used line centroid measurements to extract velocities 
from the shell lines, and these concluded that the shell is rotating. 
However, it is not clear from Figure 4 if rotation is present. The red edge of 
the SiII and FeII absorption features remains more-or-less fixed 
in position with phase, although changes in the blue half of these lines 
might at first glance be interpreted as the result of velocity variations. 
Nevertheless, the variable nature of the shell 
line profiles raises the possibility that velocities obtained from 
line centroids may be skewed by changes in line structure with 
phase. Indeed, the sub-structures that are seen in 
Figure 4 indicate that line centroids that rely on the entire absorption 
component will not probe systematic shell velocities alone, but may 
instead be sensitive to transient line profile structures that may 
not originate in the shell.

	Two sets of shell velocity measurements are 
discussed in the following sub-sections. Both sets of velocities were 
measured from phase-binned spectra that have been 
normalized to the continuum. The first set of radial velocities were 
measured via cross-correlation with a template spectrum in the wavelength 
interval 6335 -- 6465\AA\ . This wavelength interval, indicated in Figure 1, 
samples several deep shell lines, including the SiII 6347 resonance 
line. Cross-correlation techniques have the merit of yielding velocities in 
a consistent way that multiplexes the signal over a wide wavelength range. 
Of course, the resulting velocities are prone to biases introduced 
by transient features in line profiles given that the entire line contributes 
to the cross-correlation function. However, the peak of the correlation 
function is dominated by the deepest parts of the strongest lines, 
where contamination from transients is likely to be least important.

	It is evident from Figure 4 that the deepest parts of shell lines 
are less susceptible to sub-structures than the shallower components. 
Therefore, a second set of velocities were obtained from the deepest parts 
of two Si and Fe lines. A comparison of velocities measured from Si and Fe 
lines provides a means of assessing if the emission component that is 
present in the shoulders of the Fe lines (e.g. Figure 4), but is 
weaker in the Si lines, skews shell velocity measurements. 

\subsubsection{Cross-correlation velocities}

	Cross-correlation velocities were measured in an iterative manner. 
An initial set of velocities were obtained 
using the Procyon spectral atlas spectrum (Griffin \& Griffin 1979) 
as the template. The resulting cross-correlation functions 
have a clear peak from which velocities can be measured.
However, the V367 Cyg spectrum is dominated by shell features, and not lines 
that form in a stellar photosphere. Therefore, 
the initial set of velocities were used to shift the mean phase-binned 
spectra into the rest frame. The alignment of features was checked, and a 
revised template for cross-correlation that is representative of the shell 
was then constructed by stacking the rest frame spectra 
and taking the median light level at each shifted pixel 
location. Taking the median signal at each pixel suppresses features that 
do not move in sync with the shell. In Section 4.2 it is shown that the 
median spectrum constructed in this manner faithfully tracks the shell 
absorption component at all orbital phases.

	A refined set of velocities was then obtained by 
using the median spectrum as a template for the shell.
It is not surprising that the use of the median 
shell spectrum as a template greatly increased the semblance 
of the correlation function. Differences of up to $\sim 3$km/sec were 
found between the velocities obtained from the Procyon and shell 
templates, although the mean difference between the 
two sets of measurements is negligible, amounting to $0.1 
\pm 0.8$ km/sec, where the quoted uncertainty 
is the $1\sigma$ standard error in the mean. The 
velocity curve obtained with the median shell template is shown in Figure 5. 

\begin{figure}
\figurenum{5}
\epsscale{0.8}
\plotone{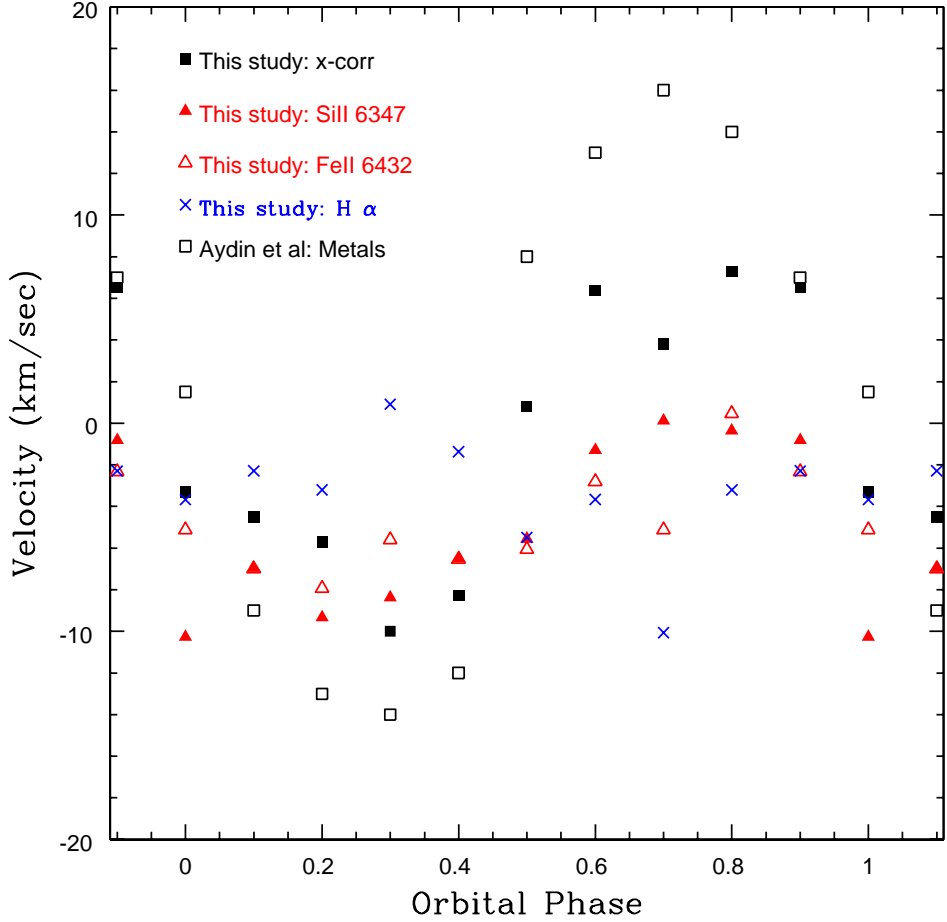}
\caption{Velocities obtained from shell lines and H$\alpha$. The filled black 
squares are velocities measured in the wavelength 
interval 6335 to 6465\AA\ using the iterative 
cross-correlation procedure that is described in the text. The open black 
squares are from the velocity curve based on metallic lines at blue 
wavelengths shown in Figure 4a of Aydin et al. (1978). The filled and open red 
triangles are velocities measured from the lower portions of the SiII 6347 
(filled) and FeII 6432 (unfilled) absorption feaures, 
while the blue crosses indicate velocities measured in 
the deepest parts of H$\alpha$ absorption. The estimated uncertainty 
in our velocities is $\pm 1$ km/sec. The velocity curves 
have different $\gamma$ values, and this 
is attributed to contamination from the donor star spectrum 
in the upper half of the metallic shell absorption 
lines. The velocities obtained from SiII 6347 and FeII 6432 
likely provide the most reliable estimate of the shell motion as 
they use the part of the line profile that is least affected by 
the donor star spectrum. Still, even those velocities may be skewed 
at some low level by sub-structure in the line profile. 
The H$\alpha$ velocities suggest motion in an opposite 
direction to that of the shell lines. The velocities 
used to construct this figure are tabulated in the Appendix.}
\end{figure}

	The velocity curve obtained from the cross-correlation measurements 
is roughly sinusoidal with an amplitude of $\sim 17$ km/sec and $\pm 1 - 2$ 
km/sec jitter. Aydin et al. (1978) measured shell velocities from TiII, FeII, 
and CrII lines at blue wavelengths, and in Figure 5 the mean of these 
are compared with the cross-correlation velocities. The 
Aydin et al. (1978) velocities were extracted from the smooth 
curve shown in their Figure 4a. Aydin et al. (1978) 
state that their velocities were measured 
from `the deepest part of the lines, rather than midway between the wings'. 
The amplitude of velocity variations obtained from the cross-correlation 
measurements is much smaller than that defined by the Aydin et al. velocities. 

\subsubsection{SiII 6347 and FeII 6432 velocities}

	A second set of velocities were measured using the 
deepest parts of the SiII 6347 and FeII 6432 lines. 
These measurements were restricted to the bottom quarter of each 
line, as this is where the line profiles at most orbital phases 
are roughly symmetric, although the SiII line near phase 0.0 is a possible 
exception. Shell velocities measured from the deepest parts of these lines 
should then be less susceptible to distortions in the profiles than 
velocities determined from the entire line.

	The velocities measured from the SiII and FeII lines are 
shown as red triangles in Figure 5. The velocity curves defined by these 
lines are more-or-less sinusoidal, with an amplitude of $\sim 6$ km/sec. This 
is considerably smaller than the amplitudes defined by the Aydin et al. (1978) 
and cross-correlation velocities shown in Figure 5. Moreover, with the 
exception of phase 0.0, where the SiII 6347 line in Figure 4 is distorted and 
the FeII 6432 is not, there is reasonable agreement between the velocities 
made from these two lines: once again, measurements made from the resonance 
SiII 6347 lines tend not to be exceptional. This agreement also suggests that 
the emission that is most pronounced in the shoulders of the FeII lines 
does not skew the velocity measurements. 

\subsubsection{Has shell rotation been detected?}

	We have presented evidence that distortions in the profiles of shell 
lines have likely biased shell velocity measurements in previous studies. 
Heiser (1961) reviews previous measurements that span a time baseline of many 
decades, and notes that the $\gamma$ velocity of the shell velocity curve 
varies over timescales of a few years, spanning a $\sim 30$ km/sec range. 
However, the results in Figure 5 indicate that the $\gamma$ velocity of the 
shell is susceptible to the procedure used to measure velocities. 
The $\gamma$ value from the cross-correlation measurements 
is $\sim 0$ km/sec, while that obtained from the SiII and FeII measurements 
is $\sim -4$ km/sec. The Aydin et al. (1978) measurements in Figure 5 
yield $\sim 2$ km/sec. We suspect that the 
variations in $\gamma$ found in previous studies are due to differences 
in the method used to estimate the radial velocities of shell lines, 
rather than a physical shift in the mean shell velocity.

	We suspect that the velocities made from SiII 6347 
and FeII 6432 more faithfully track the true kinematics 
of the shell than the other measurements shown in Figure 5 as they are 
based on the parts of the profiles that appear to be least affected 
by distortions. Even then, the residual contamination evident in Figure 4 even 
near the line centers may still skew the results. Hence, the amplitude of the 
velocity measurements attributed to the shell 
may be even smaller than that found from these lines. 
A constant shell velocity with phase would be consistent with the 
azimuthal uniformity of the shell that is required to remove the 
shell component from the spectra at all orbital phases (Section 4.2).

\subsection{Stability of the Shell Spectrum with Orbital Phase}

	Insights into the origins of the complex structure in the shell lines 
can be gleaned by examining the differences between 
individual mean phase-binned spectra and the median shell 
spectrum. Changes in the shape of the line profile 
with orbital phase could signal a non-uniform distribution of 
circumsystem material and/or contributions from sources that 
are not physically related to the shell. Differenced spectra, 
in the sense of the mean spectrum at a given phase minus the 
median shell spectrum at wavelengths centered on SiII 6347, SiII 6371 
and FeII 6432 are shown in Figure 6. The median shell spectrum in each 
wavelength interval is also shown. 

\begin{figure}
\figurenum{6}
\plotone{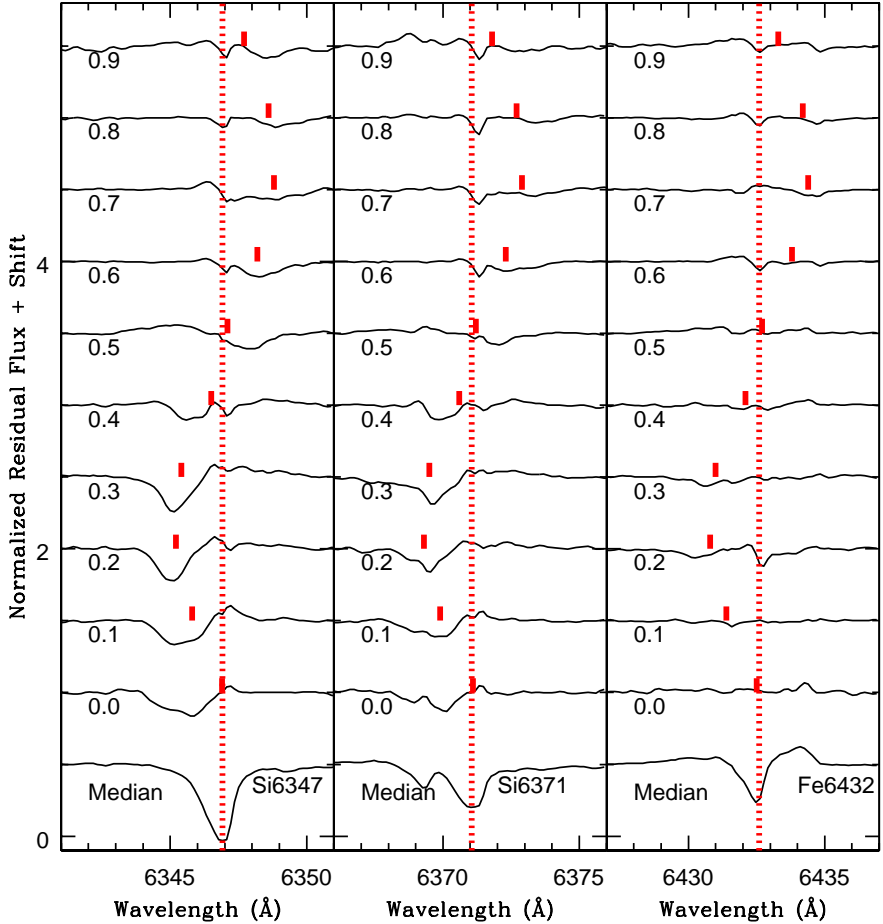}
\caption{Differences between the median shell spectrum 
and the mean phase-binned spectra near SiII 6347, SiII 6371 and FeII 6432. 
The median shell spectrum is shown at the bottom of each panel. 
The dotted line marks the point of deepest absorption in each shell line, while 
the expected locations of the shell lines if they 
were in the spectrum of the donor star, assuming the 
composite velocity curve in Figure 6 of Tarasov \& Bergyugin (1998), 
are indicated in red. There is a residual absorption feature that is deepest 
at phases 0.2 and 0.3 that follows a serpentine-like 
propogation with phase. The $\sim 3$\AA\ throw 
of this feature in wavelength between the two quadrature points is consistent 
with that expected for the orbital motion of the donor star. SiII 
lines are among the deepest metallic signatures in the spectra of A-type 
stars at these wavelengths, whereas FeII lines tend to be weaker than the Si 
lines at those spectral types.}
\end{figure}

	The differenced spectra in Figure 6 indicate that 
the dominant shell absorption feature at the rest wavelength 
of all three lines has been removed to within the few percent level
at all orbital phases. Given that the phase-averaged spectra are 
constructed from observations that span a range of epochs, then the stability 
of each shell absorption line is consistent with a radially uniform 
distribution of material throughout the shell over many 
orbital cycles. This is consistent with hydrodynamical 
simulations of the mass outflow from model binaries that 
predict a shell that is smoothed by shocks at large radii (e.g. 
Bermudez-Bustamante et al. 2020).

	While the main components of each shell line appear to have a stable 
depth over many orbital cycles, variations in the residual spectra with orbital 
phase are evident, and these are most pronounced in the SiII 6347 
and SiII 6371 lines. Similar residuals are seen in the FeII 6432 line, 
although with a smaller depth than in the Si 
lines. The behaviour of the SiII 6347 resonance line is similar 
to that of the other metal lines, although in the next Section 
it is shown that small-scale sub-structures in the SiII 6347 lines may 
be present. The origin of the dominant feature in the residual 
spectra of each shell line is the subject of the next Section.

\section{THE SPECTRUM OF THE DONOR STAR}

	Given that lines from the accreting star have not yet been 
detected, then only the spectrum of the donor star has been explored to date. 
Based on Mg II 4481 and other lines that may contain spectroscopic signatures 
of the donor star, Heiser (1961) assigned that star a spectral type 
A2, while Aydin et al. (1978) assign it type A5I-II. Estimates of the 
temperature of the donor star have ranged from 10000 K (Zola \& Ogloza 2001) 
to 13450 K (Pavolvski et al. 1992). The rapid rotation 
of stars in CBSs complicates the relation between 
effective temperature and spectral-type (e.g. Yakut \& Eggleton 2005). 
Large-scale spot activity on the donor might also frustrate efforts to 
determine its spectral-type and effective temperature.

\subsection{Extracting the Donor Star Spectrum}

	There is a compelling motivation to identify lines in the 
donor star spectrum over a range of wavelengths, as these can be used to 
eaxmine basic characteristics such as the rate of rotation, chemical 
content, and effective temperature. Lines in 
the donor star spectrum have been identified 
at blue wavelengths (e.g. Heiser 1961; Aydin et al. 1978), and 
the most extensively studied of these is MgII 4481. An interesting finding 
is that spectra presented by Heiser (1961) and Aydin et al. (1978) show 
that the character of MgII 4481 changes with orbital phase, 
in the sense that it is deep and sharp near phase 0.25, but is weaker 
and broader near phase 0.75.

	Lines in the donor star spectrum are detected in our data. 
There is prominent residual absorption 
in Figure 6 that weaves its way with wavelength through the 
SiII and FeII residuals. The change in the wavelength of these features 
from phase 0.2 to 0.6 is $\sim 3$\AA\ , corresponding to a $\sim 140$ 
km/sec velocity variation. This is consistent with the amplitude of 
the donor star radial velocity curve measured from MgII 4481 (e.g. Figure 6 
of Tarsov \& Berdyugin 1998). The vertical red lines in Figure 6 mark 
the expected location of the Si and Fe lines with orbital phase, working on 
the assumption that they originate in the photosphere of the donor star and 
that they follow the velocity curve of Tarasov \& Berdyugin (1998). 
The absorption features in the residual spectra tend to have wavelengths that 
are consistent with an origin in the donor star spectrum.
Moreover, the depths of the residual features in Figure 6 
also vary with orbital phase, in a manner that is similar to 
that seen in MgII 4481.

	Cycle-to-cycle changes in the mass transfer rate and/or 
the rate of mass deposition onto the accreting star 
could affect the outflow of material from the system, thereby producing 
structure in this material and signatures in shell 
absorption lines. Sub-structure in the mass flow might be most 
obvious in a resonance line, like SiII 6347, and the residual spectra 
of this feature do show subtle orbit-to-orbit 
variations. Evidence for this is shown in Figure 7, where residual spectra 
of SiII 6347 from the three epochs that are covered in the mean  
spectrum for phase bin 0.1 are shown. The spectra span a 4 month 
baseline, from May 2021 to September 2021.

\begin{figure}
\figurenum{7}
\epsscale{0.8}
\plotone{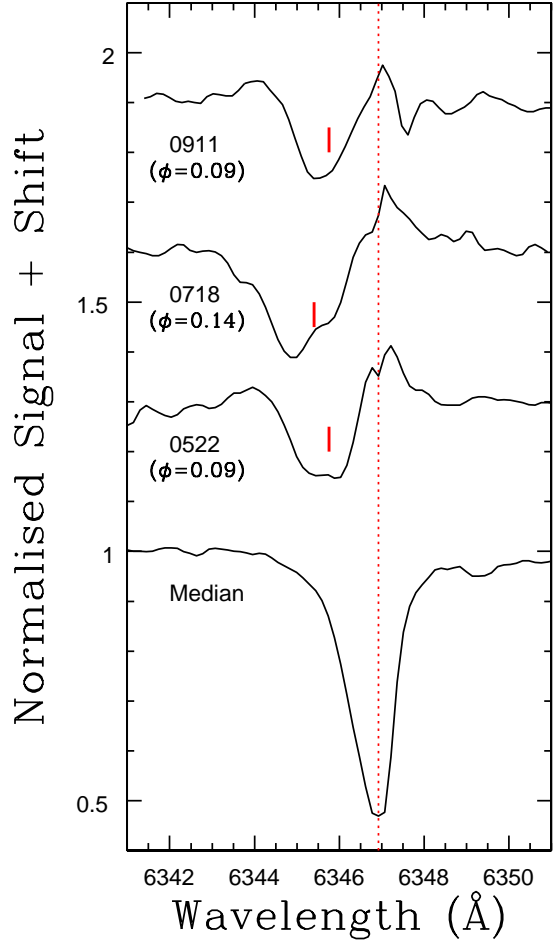}
\caption{Residual structure in the SiII 6347 feature on 
three different nights that cover the phase 
interval 0.05 - 0.15. The orbital phase ($\phi$) for each spectrum is shown. 
The residual spectra were constructed using the same procedure used to make 
Figure 6, and the dotted red line marks the wavelength where the shell line 
is deepest. The expected wavelength of SiII 6347 in the donor 
star spectrum at each orbital phase is indicated in red. The residual 
spectra on all three nights show broad absorption near 6345\AA\ , and this 
is attributed to the donor star spectrum. There is evidence of a second feature 
near 6346\AA\ that appears to weaken progressively with time over the 
range of dates sampled by the spectra. Thus, while much of the residual 
structure can be attributed to the donor star absorption spectrum, 
there may be transient components that originate in other sources.}
\end{figure}

	There is prominent absorption centered near 6345\AA\  in the 
residuals, and the overall depth of this feature does not change markedly
with time. The expected location of SiII 6347 in the donor star spectrum if it 
were to follow the velocity curve shown in Figure 6 of Tarasov \& Berdyugin 
(1998) is indicated by the solid vertical red line above each spectrum. 
The deepest feature in the residual spectra 
falls a few tenths of an \AA\ blueward of the expected line position. 
This mis-match in the predicted and observed 
residual absorption feature is likely not due to 
an error in the velocity curve adopted for the donor star. 
It can be seen from Figure 6 that an offset between the expected and 
observed location of the residual absorption is restricted to the SiII 
6347 line at phase 0.1, and increasing the overall amplitude of the velocity 
curve will result in a mis-match between the expected and observed 
line wavelengths at other phases. We thus suspect that not all of the 
sub-structure within the SiII 6347 line is associated with the spectrum of the
donor star; while the central wavelengths of the 
residuals in Figure 6 tend to vary with orbital phase in a manner that is 
consistent with them tracking the motion of that star, 
signatures from other components might also be present.

	A donor star spectrum can be constructed by shifting 
the phase-binned residual spectra to the rest frame 
after adopting a radial velocity curve for that star. 
The velocity-corrected spectra can then be combined to suppress 
artifacts in the residuals that are not due to the donor star. 
This basic procedure was adopted here, and 
donor star spectra were constructed for phases 0 - 0.5 and 0.5 - 1.0 
given evidence that the depths of absorption features vary with phase. 

	The donor star spectra are shown in Figure 8, 
along with the mean spectrum that was constructed from all phases. 
Based on a model that assumes that the accreting star is enshrouded in a disk, 
Zola \& Ogloza (2001) find that the donor star contributes 80\% of the total 
system light at red wavelengths. Given the evidence of an accretion disk in 
the system, we adopt this fractional light contribution from the donor star. 
Therefore, the light levels of the donor star spectra 
in Figure 8 have been elevated by a factor of 1/0.8 from their extracted values 
to display the strengths of features expected in the spectrum of a single star. 

\begin{figure}
\figurenum{8}
\plotone{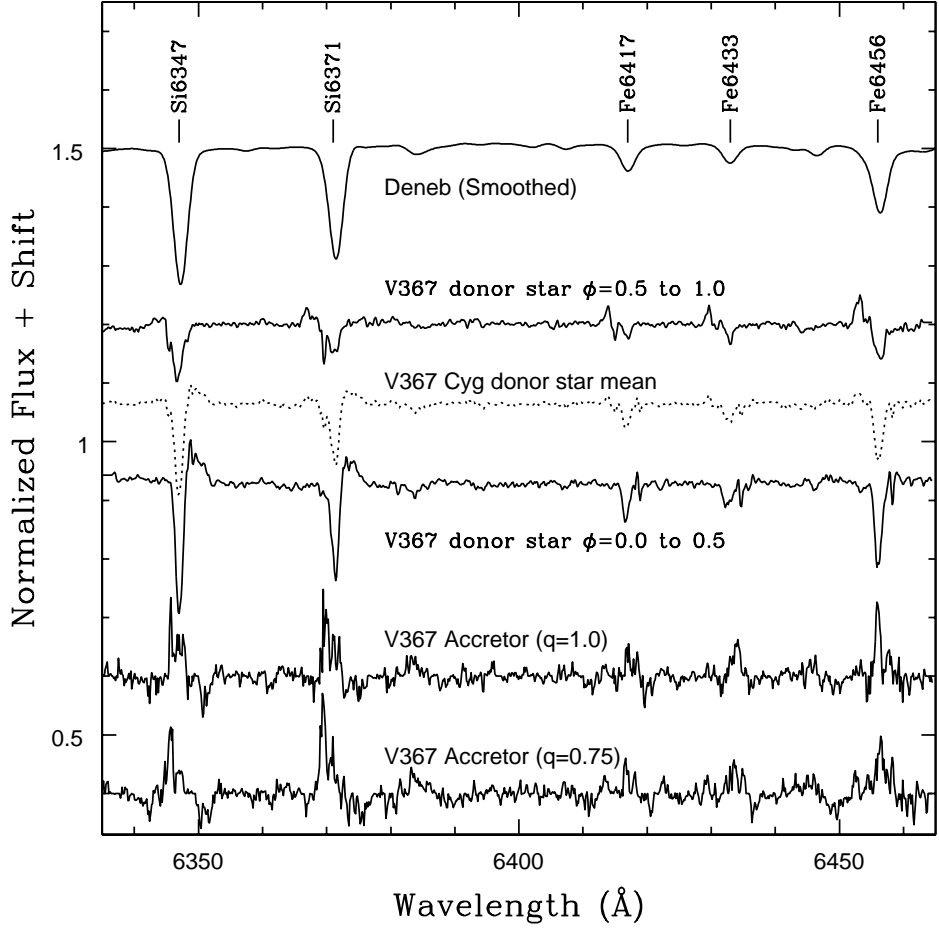}
\caption{Spectra of -- from top to bottom -- Deneb, the donor star 
at different phases ($\phi$), and the accretion disk. The Deneb spectrum has 
been smoothed to match the widths of lines in the mean donor spectrum. 
Donor star spectra in two phase intervals 
(0.0 to 0.5, and 0.5 to 1.0) are shown, as is 
the mean spectrum over all orbital phases (plotted as a dotted line). 
The spectrum of the donor star between phases 
0.0 and 0.5 is similar to that of Deneb, suggesting that 
the donor star has features that are consistent with those of a 
mid-A giant/supergiant in this phase interval. The accretion disk 
spectra were constructed for two system mass ratios (q), and 
emission features of SiII and FeII are apparent. The strengths 
of the emission features are lower limits to their actual values given 
the procedure used to construct these spectra (see text). 
The donor star and accretor spectra have been scaled to produce lines 
with strengths that are appropriate for a single object, based on the 
relative luminosities listed in Table 2 of Zola and Ogloza (2001) 
in the R filter.}
\end{figure}

	SiII and FeII absorption lines are seen in the donor star 
spectra. SiII 6347 and SiII 6371 are among the most prominent 
metallic lines in the spectra of A-type stars at these wavelengths, and 
the presence of these lines is perhaps not surprising given 
previous spectral-type estimates for the donor. 
In Section 5.2 it is shown that the depths of these lines 
are consistent with a mid-A spectral type. 

	The SiII and FeII lines are weaker and broader near orbital phase 0.75 
than those near phase 0.25. The donor star spectra in 
the phase intervals centered on these phases also differ in the sense that 
there are emission components in the shoulders of the absorption lines that 
are shifted in different directions from the line centers. The 
nature of these shifts is consistent with the orbital motion of the accreting 
star. These are likely artifacts of emission lines that originate in the 
accretion disk around that star, and these are seen throughout the residual 
spectra in Figure 6. 

	The variation in the depths of SiII and FeII lines in the 
donor star spectrum with orbital phase is reminiscent of the 
behaviour of MgII 4481. Given that this similarity is based on 
Mg II spectra that were recorded at much earlier epochs, then 
the source of any phase dependency in the donor star spectrum has likely 
been stable for at least $\sim 50$ years, or almost 1000 orbital cycles. 
Such stability suggests that the donor star is tidally locked 
with the motion of the binary system at the present day, and this is 
consistent with the rotation velocity estimated from the line widths and some 
system geometries found from light curve modelling (see below). 
That the spectra differ at the quadrature points leads us to suspect 
that the phase-dependence of line strength is due to spot activity that 
covers much of a stellar hemisphere, such that it dominates the light over a 
wide range of orbital orientations. 

	The rotation velocity is an important parameter for stars in binary 
systems as rotation-induced mixing affects the internal 
structure of a star, and hence the pace of evolution and the properties 
of the final remnant (e.g. Taormina et al. 
2020). If assumptions are made about tidal synchronization 
and geometric properties such as the Roche lobe fill 
factor then the rotational velocity can also be used to test 
system parameters. The widths of the SiII 6347 and SiII 6371 lines in 
the residual spectrum were measured in the donor star 
spectrum at phases 0.2 and 0.3, where these features are deepest and hence 
best defined. The line widths are consistent with $v sini \sim 54 
\pm 2$ km/sec FWHM, where the uncertainty is the $1\sigma$ formal error in the 
mean. This is consistent with previous rotation velocity estimates for this 
star (e.g. Pavlovski et al. 1992). Moreover, this rotational 
velocity is consistent with a tidally locked donor star with a 
size of 21R$_{\odot}$, as found by Zola \& Ogloza (2001) from light curve 
models that include an accretion disk.

\subsubsection{Comparisons with Deneb}

	Previous studies suggest that the donor 
star spectrum should be similar to that of an 
evolved A star, such as Deneb (spectral type A2Ia). Therefore, 
spectra of Deneb that were recorded as part of program 
DAO122-2015C4 (PI: Yang) were retrieved from the CADC archive and reduced using 
the same steps discussed in Section 2. These spectra 
were recorded with the DAO 1.2 meter telescope 
with the same instrumental configuration used to observe V367 Cyg.

	The processed Deneb spectrum was smoothed with a Gaussian to match 
the widths of the lines in the mean donor spectrum, and the result is shown in 
Figure 8. The SiII lines in the mean donor spectrum are weaker than those 
in the Deneb spectrum, although there is better agreement with the 
donor spectrum between orbital phases 0 and 
0.5. Similar conclusions hold for the FeII lines. 
Between phases 0.5 and 1.0 the Si and Fe 
lines in the donor star spectrum are weaker than in Deneb. 
While the agreement with Deneb is not perfect, the comparisons in Figure 8 
still indicate that the donor spectrum is consistent with that of an 
early to mid-A giant/supergiant. Extracting a donor spectrum 
at other wavelengths would help to better define the spectral-type 
of the donor.

\section{EMISSION FEATURES}

\subsection{Si and Fe Emission Associated with the Accretor}

	V367 Cyg is a single line spectroscopic binary, and this frustrates 
efforts to obtain reliable component masses via traditional kinematic means.
However, if lines from the accretion disk can be recovered then they can be 
used to trace the motion of the accreting star. Emission lines from 
accretion disks have been used to probe the orbital motions 
of white dwarfs in recurrent novae (e.g. Sahman et al. 2013). 
Tarasov \& Berdyugin (1998) estimate stellar masses in 
V367 Cyg using a purported detection of HeI 6678 
that they associate with the accretion disk. 

	The detection of spectroscopic 
signatures from the accretion disk is not an unreasonable expectation for 
V367 Cyg, as the light curve solution presented by Zola \& 
Ogloza (2001), in which the secondary star is modelled as 
a disk, predicts that $\sim 20\%$ of the broad-band system light comes from 
that structure. The emission features in the shoulders of the 
absorption lines propogate in the opposite direction to features 
that originate in the donor star spectrum (e.g. Figure 6), 
leading us to suspect that they form in the disk around the accreting star. 

	A complicating factor when attempting to 
reconstruct the emission line spectrum is that the 
central regions of these lines are obscured by shell absorption. 
Still, if a system mass ratio is assumed then residual spectra at 
various phases can be placed into the rest frame of the accreting star, and the 
results combined to search for spectroscopic signatures associated with that 
star and/or the surrounding disk. While this will likely yield spectra with 
compromised line strengths due to shell absorption lines, it may still provide 
clues into the nature of the emission -- for example, are there 
spectroscopic similarities with other early-type stars that 
have accretion disks?.

	The results of combining the residual spectra 
if system mass ratios of unity and 0.75 are assumed are shown at the 
bottom of Figure 8. These spectra were constructed by taking the median 
signal at each aligned pixel after shifting to the rest frame using 
velocities for the accreting star calculated from 
the Tarasov \& Berdyugin (1998) radial velocity curve 
for the donor star. The spectra have been scaled upwards 
using the R-band luminosity for the disk listed in Table 2 of Zola 
\& Ogloza (2001). A spectrum was also constructed for a 
mass ratio of 0.5, and the emission features in that spectrum are much 
weaker than in the spectra that assume higher 
mass ratios. This suggests that the system mass ratio is $> 0.5$.

	The spectra for mass ratios of 0.75 and 1 are very similar. 
Given the presence of prominent emission features in the 
observed spectra that move in sync with the 
expected motion of the accreting star then it is not surprising that line 
emission is present in the combined spectra. However, 
the accretion spectra in Figure 8 are noisy, and this is contrary 
to what might be expected given the prominent emission lines in the 
unphased spectra. We emphasize that the strengths of the emission 
lines in Figure 8 are lower limits to their actual values because the central 
regions of the emission lines overlap with the shell absorption lines, with the 
result that much of the signal in the central part of 
the emission line is lost when constructing 
the residual spectra. This produces the clipped appearance of the emission 
lines in Figure 8. The absorption features from 
the donor star are also broadened by rotation, and 
at least some level of coherence in the donor star spectrum is retained. 
This will also affect the strengths and widths of accretion disk features.

	An examination of the excitation mechanisms that power 
the emission is frustrated by the problems recovering reliable line strengths 
and widths. Still, we note that Fe and Si emission lines are seen in the 
spectra of other types of objects in which early-type stars are surrounded by 
an accretion disk, including Herbig AeBe stars (e.g. Hernandez et al. 2004) and 
classical Be stars (e.g. Mathew \& Subramaniam 2011). This similarity 
extends to other features in the V367 Cyg spectra, such as 
H$\alpha$ and [OI] 6364, and these similarities are discussed 
in the following sections.
 
\subsection{H$\alpha$}

	Tarasov \& Berdyugin (1998) discuss the behaviour of H$\alpha$ 
in V367 Cyg, and find that there is prominent emission with 
deep central absorption. The characteristics of both the emission and 
absorption components were found to remain constant with orbital phase, 
although the Tarasov \& Berdyugin spectra have limited phase coverage, with 
three of their four profiles covering orbital phases between 0.6 and 0.8. The 
spectra recorded for this study track H$\alpha$ over a full range of 
phases.

	Mean phase-binned spectra centered on H$\alpha$ are shown in 
the left hand panel of Figure 9. Previous studies have 
identified absorption from higher order Balmer lines that is believed to 
originate in the donor spectrum. The spectrum of the A-type supergiant 
Deneb is shown in Figure 9. To the extent that Deneb is a good match 
to the spectrum of the donor star then it is clear that the donor 
makes only a modest contribution to the observed H$\alpha$ absorption. 

\begin{figure}
\figurenum{9}
\plotone{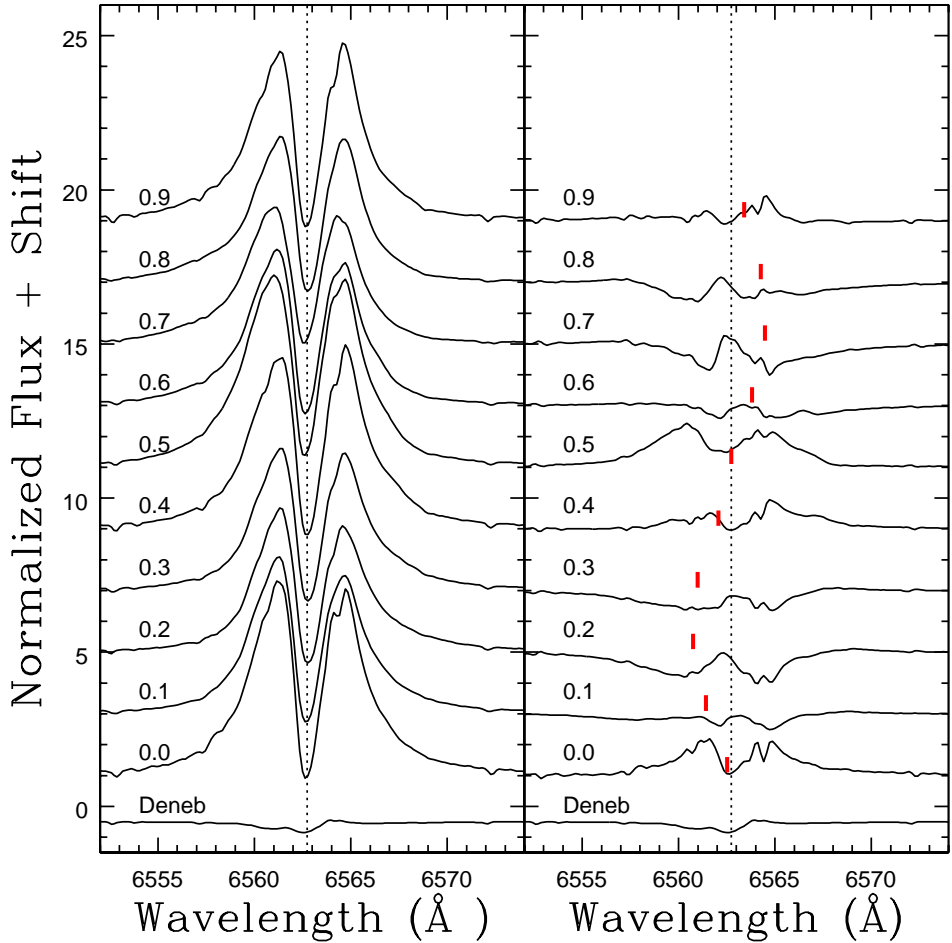}
\caption{Left hand panel: Phase-binned spectra centered on 
H$\alpha$. The spectrum of Deneb is shown at the bottom of the panel, 
and it is evident that H$\alpha$ absorption from an A supergiant spectrum like 
that of Deneb makes only a modest contribution to the 
total H$\alpha$ absorption. The dotted line indicates 
the mean midpoint of H$\alpha$ absorption, as measured 
from the lower portions of the line profiles. Modest changes in the 
central wavelength of the absorption component and in the shape of 
the emission component are apparent. Right hand panel: Differences 
between the composite mean H$\alpha$ spectrum constructed by averaging the 
spectra at all ten phases and the phase-binned spectra. The residuals highlight 
changes in the amplitude of the emission and absorption features 
with orbital phase. The expected location of 
H$\alpha$ absorption in the donor spectrum is indicated with 
red lines, and any signatures of this feature are swamped by 
noise in the residuals.}
\end{figure}

	When interpreted in the scheme described by Reipurth et 
al. (1996) for Herbig AeBe stars, the H$\alpha$ 
profiles would be classified as Type II, which are the most common type 
of profile (e.g. Viera et al. 2003). H$\alpha$ emission in V367 Cyg is 
likely circumsystem in origin, forming in material that has been 
ejected from the system. This is consistent with the modest shifts in 
wavelength with orbital phase that are evident in Figure 9. 

\subsubsection{H$\alpha$ variations with phase}

	There appears to be a slight change in the central wavelength 
of H$\alpha$ with orbital phase, suggesting 
a variation in radial velocity. This being said, the asymmetric nature of the 
H$\alpha$ absorption profile is a source of uncertainty when measuring 
velocities. As these asymmetries appear to be smallest 
near the bottom of the H$\alpha$ absorption line then we 
have measured velocities from that part of the profile, and 
the results are compared with the donor star and shell velocities in Figure 5. 

	The amplitude of the velocity variation defined by H$\alpha$ is 
small, and a systematic variation with phase is not seen. Still, 
the velocities near phases 0.25 and 0.75 hint at an 
orbital motion in the same direction as the accreting star. 
Tarasov \& Berdyugin (1998) find evidence 
of a 15 km/sec systematic shift in the velocity of H$\alpha$ when 
compared with other features. Three of their four profiles are between 
phases 0.6 and 0.8, and this is where the largest departure from the mean 
H$\alpha$ velocity is seen in our measurements. 

	The shapes and central wavelengths of the H$\alpha$ absorption 
and emission components in Figure 9 change with orbital phase. To examine 
these changes, the spectra were aligned using the velocities measured from 
H$\alpha$ absorption discussed above, and a mean spectrum was generated. 
The mean spectrum was then subtracted from each of the 
phased-binned spectra, after they too had been aligned using 
the velocity measurements made from H$\alpha$ absorption. 
The residual spectra are shown in the right hand panel of Figure 9.

	H$\alpha$ changes with orbital phase. 
The largest deviations from the mean spectrum occur at the eclipses, 
in the sense that the absorption and emission components have 
the largest residuals at these phases. This helps to explain why variations 
with phase were not detected by Tarasov \& Berdyugin (1998), as 
none of their spectra fell within $\pm 0.05$ in phase of an eclipse. 
The variation in H$\alpha$ strength with phase in Figure 9 could be a 
geometric effect in which the binary system blocks light from some 
of the emitting gas outside of eclipse.

	We have examined if signatures of H$\alpha$ absorption from the 
donor star spectrum can be seen in the residual 
spectra. The expected wavelengths of H$\alpha$ absorption in the donor 
star spectrum at various phases are marked in the right hand column of Figure 
9. These wavelengths were calculated using the Tarasov \& Berdyugin (1998) 
composite velocity curve for the donor star. 
Absorption features are not seen at the expected wavelengths. 
That H$\alpha$ absorption from the spectrum of the donor star has not 
been detected is likely due to the broad and strong H$\alpha$ emission 
component that is present at all orbital phases, coupled with noise 
in the residuals.

\subsubsection{H$\alpha$ near secondary minimum}

	H$\alpha$ emission and absorption are stronger than average near 
secondary minimum, and this is also where there are difficulties matching 
the light curves (e.g. Zola \& Orgloza 2001). As the dominant feature in the 
red spectrum, the behaviour of H$\alpha$ at phases near secondary 
minimum may provide clues into the origin and distribution of 
light in that part of the light curve. This motivated us to 
examine individual spectra that were recorded on individual nights 
and that sample the 0.45 - 0.55 phase interval.

	There are eight spectra in our dataset that sample phases between 
0.45 and 0.55, and residual H$\alpha$ spectra for these nights 
are shown in Figure 10. The residuals are the 
result of subtracting the mean spectrum constructed from all orbital 
phases from each spectrum. The results are ordered according 
to the offset in phase from secondary minimum, 
such that those closest to secondary minimum are shown at the top, and phases 
that are progressively more removed from secondary minimum are placed towards 
the bottom of the figure. This display scheme allows the symmetry of 
H$\alpha$ about secondary minimum to be evaluated, while also enabling a 
search for variations with time.

\begin{figure}
\figurenum{10}
\plotone{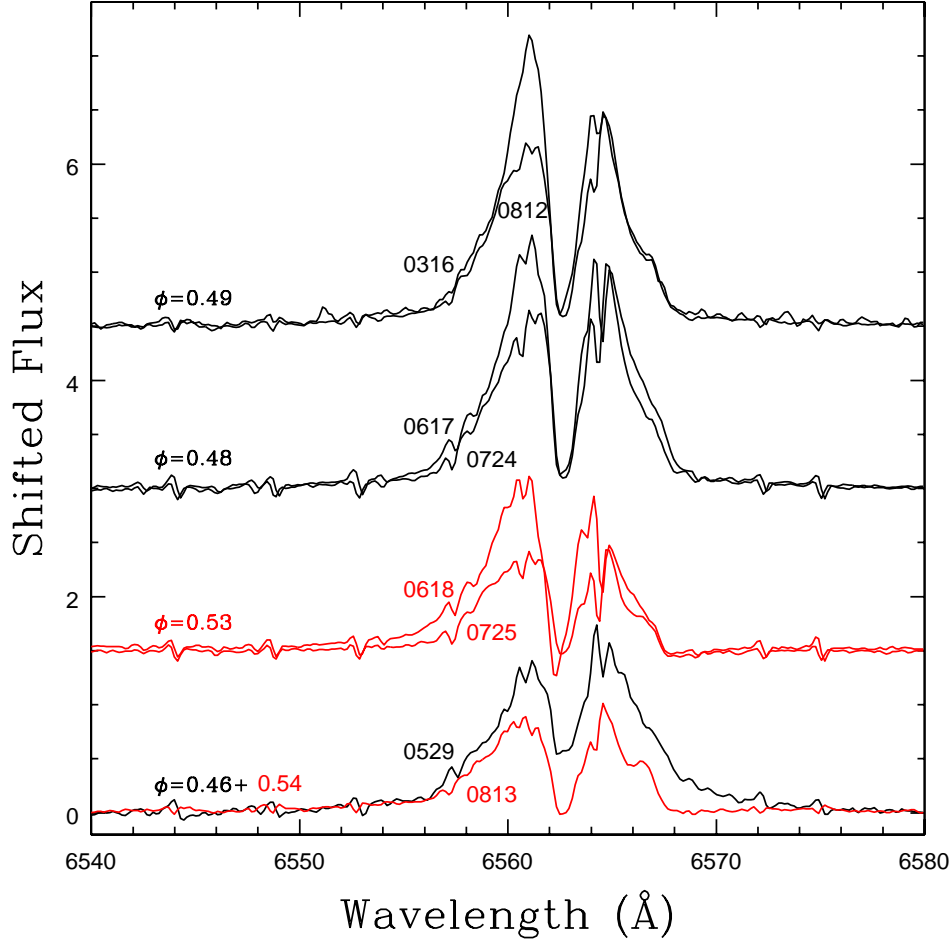}
\caption{H$\alpha$ residuals near secondary minimum 
(phase 0.5). Each curve shows the result of 
subtracting the phase-averaged mean spectrum from an individual spectrum. 
Spectra are grouped using orbital phase bins having a width of 0.01 phase 
units, and the spectra are identified by the month and day in that month 
(e.g. March 16 = 0316). The residual spectra are displayed folded about 
phase 0.5 so that the brightening and fading of H$\alpha$ can be compared 
before and after secondary minimum. The spectra at the top are closest to 
secondary minimum in terms of orbital phase, while the spectra at the 
bottom are furthest from secondary minimum. Spectra recorded with phases 
between 0.51 and 0.55 are shown in red. There are differences between 
the spectra at each phase offset from secondary minimum, and it is argued in 
the text that these are likely due to variations in H$\alpha$ strength with 
time, rather than orbital phase, in the sense that H$\alpha$ emission was 
consistently stronger during March -- June 2021. 
There is also evidence for systematically stronger H$\alpha$ 
emission towards phase 0.5, based on the strengths of spectra that were 
recorded only one day apart in June (0617 and 0618) and July (0724 and 0725).}
\end{figure}

	There are obvious differences between the spectra at phases 0.46 and 
0.54, in the sense that H$\alpha$ is weaker at phase 0.54. 
We suspect that this difference is likely due to variations in the strength 
of H$\alpha$ over timescales of a few months, rather than due to orbital 
phase. More specifically, there are significant differences between spectra 
that sample the same phase but that were recorded on different dates. These 
differences are such that H$\alpha$ emission from the system appeared 
to be stronger during March - June 2021 than during July - August of that year. 
The two spectra at phase 0.48 are separated in time by only 37 days, 
indicating that the H$\alpha$ strength varied considerably over only 
two orbital cycles. When considered in the context of an outflow model, 
such as that discussed by Deschamps et al. (2015), the variations in 
the strength of H$\alpha$ emission with time that are seen in Figure 
10 could be due to enhanced ejection rates and 
clumpy structure in the outflow, as might form if material is 
not ejected from the system at a uniform rate.

	In addition to evidence that H$\alpha$ varies with time, 
the residual spectra in Figure 10 indicate that H$\alpha$ varies in 
a systematic way with orbital phase. There are two sets of 
spectra recorded in June and July that are separated by only 
1 day ($\sim 0.05$ in orbital phase), and these show a clear tendency for 
H$\alpha$ to be stronger at phases closest to secondary minimum. 
Both time and orbital phase thus appear to affect the strength of H$\alpha$ 
emission near secondary minimum. As the H$\alpha$ flux during ingress and 
egress of secondary minium appears to be symmetric with phase, then 
the variation of H$\alpha$ with time may have 
contributed to difficulties matching secondary minium in the light curves.

\subsection{[OI] Emission}

	Broad [OI] emission with a width of $\sim 4.5$ \AA\ is present at 
all phases. [OI] 6364 will be accompanied by the much stronger [OI] line 
at 6300 \AA\ , although this transition falls just outside of the 
wavelength range covered by our spectra. Herbig AeBe stars may provide 
some insights into the origin of [OI] emission in V367 Cyg. 
[OI] 6364 is seen in the spectra of many Herbig AeBe objects 
(e.g. Bohm \& Catala 1994), where there is evidence that it is 
associated with an optically thick disk and outflow (e.g. Corcoran \& Ray 
1997, Viera et al. 2003, Acke et al. 2005).
[OI] 6364 emission in these stars is often accompanied by Si and 
Fe emission lines, such as are seen in the shoulders of the shell 
absorption lines that have been the focus of much of this paper. 
[SII] 6716 and 6731 emission can also be present. 
Figure 3 shows hints that [SII] 6731 might be present 
at some phases, although it is very weak. Efforts to detect [SII] 6716 
emission in Figure 3 are frustrated by the FeII 6718 shell line. 

	While the spectra of most Herbig AeBe stars have a pronounced [OI] 
emission peak (Bohm \& Catala 1994), this is not the case here. Rather, the 
broad nature of [OI] 6364 in the V367 Cyg spectrum suggests that it originates 
in an environment with a $\pm 100$ km/sec velocity dispersion. The shape 
of the [OI] 6364 profile also changes with orbital phase. 
This is demonstrated in Figure 11, where phase-binned spectra are 
shown in the left hand panel. The results of subtracting 
the phase-averaged spectrum are shown in the right hand panel of this figure. 
Contamination from SiII 6371 emission blueward of the main shell absorption 
lines can be seen near 6367.5\AA\ in the spectra at phases 0.0 and 0.1 
(see also Figure 1). 

\begin{figure}
\figurenum{11}
\plotone{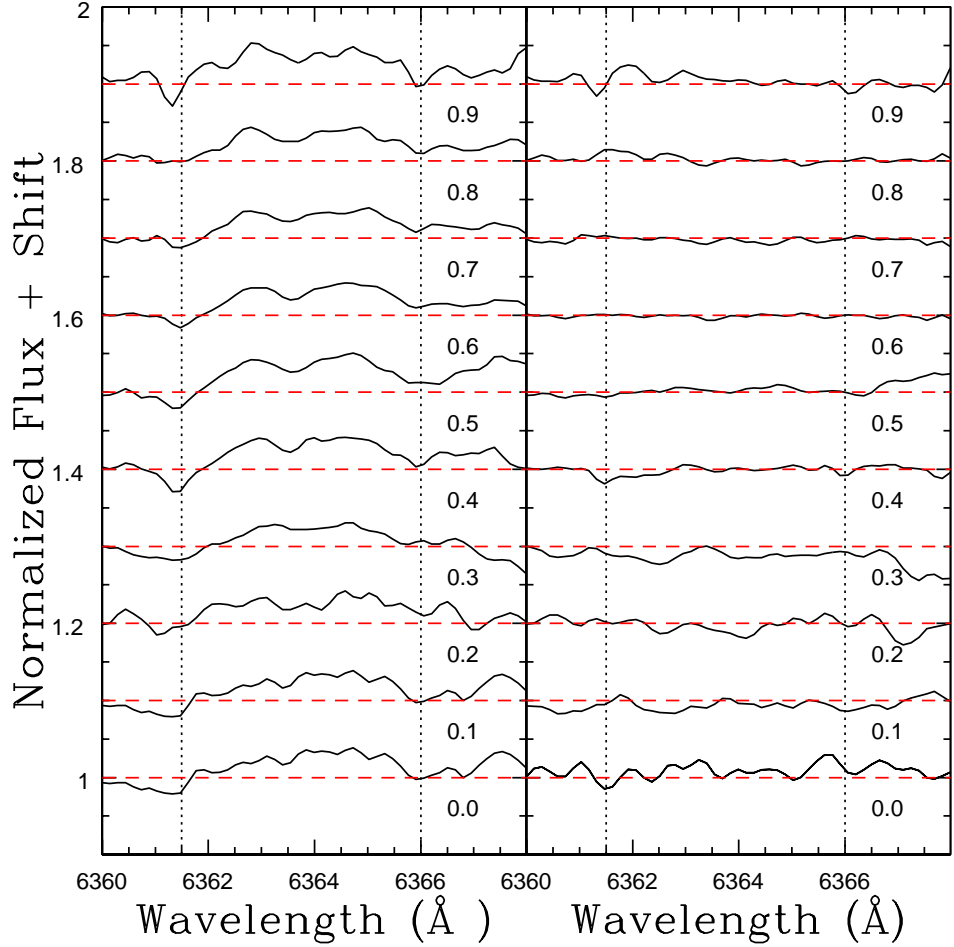}
\caption{Left hand panel: Mean phase-binned spectra 
of [OI] 6364. The continuum level for each spectrum is indicated 
with a red dashed line, while the dotted lines mark the approximate 
blue and red boundaries of the emission.  Emission 
in the blue part of the [OI] feature is weakest at phases 0.0 and 0.1. 
The component in the blueward part of the line grows in strength towards 
progressively later orbital phases, while the strength of the redward signal 
stays roughly constant. Right hand panel: Differences between a composite 
mean spectrum constructed by averaging the spectra at all 
ten phases and individual phase-binned spectra.}
\end{figure}

	It has been suggested that [OI] emission 
from Herbig AeBe stars originates in a circumstellar disk that is 
undergoing Keplerian rotation (e.g. Acke et al. 2005).
The structure that is seen in the [OI] 6364 emission 
suggests that there may be multiple emission sources, although there is 
evidence that a sizeable fraction of this emission may come from the 
accretion disk. In particular, there is a tendency for 
the emission at the short wavelength portion of the feature to 
strengthen towards later phases, while the emission at redder wavelengths 
stays more-or-less constant. This is reminiscent of 
the behaviour of the SiII and FeII emission components seen in the 
shoulders of the shell lines. We have not attempted to measure velocities 
from [OI] 6364 given that this is a broad, complex feature. However, 
we note that [OI] emission in Figure 11 spans the wavelength range 
that would be expected if the feature originated in a disk that moved 
with the secondary star if the system had a mass ratio near unity.

	Blue-shifted [OI] 6300 emission is a common feature in Herbig AeBe 
stars. The most extreme cases are associated with stellar jets and outflows 
in which light from corresponding red-shifted emission is blocked 
by the disk (e.g. Corcoran \& Ray 1997, Acke et al. 2005).
A component of the [OI] emission from V367 Cyg 
might also originate in an outflow, that presumably originates from the 
hot spot or the outer Lagrangian points. If present, then such a component 
would likely contribute to the red side of the [OI] emission, as that 
part of the line appears to be stable with phase. An examination of [OI] 6300 
will provide additional insights into the origins of [OI] emission from 
V367 Cyg.

\subsection{HeI 6678}

	Given that they are experiencing rapid mass transfer, 
W Ser systems are expected to contain a hot spot where the mass 
stream strikes the accretion disk. If it has a high enough effective 
temperature then this hot spot has the potential to power some of 
the unusual spectroscopic features associated with these systems, and also 
explain some of their unusual photometric properties, including features 
in the light curve that are not attributed to the component stars. Table 1 of 
van Rensbergen et al. (2011) provides a compilation of hot 
spot temperatures on the accretion disks of W Ser systems, and in 
many cases the temperatures are high enough to power HeI emission. 
While not expected to originate directly from the photospheres of 
the stars in V367 Cyg, a search for HeI 6678 emission is then warranted.

	Tarasov \& Berdyugin (1998) discuss efforts to detect HeI 6678 
emission from V367 Cyg. They identify a feature 
with radial velocities that are consistent with those 
expected for the accreting star and its accompanying accretion 
disk, and identify it as HeI 6678. However, HeI 6678 is at a 
wavelength where there are also shell absorption 
lines (e.g. Figure 3), and this lead Zola \& Ogloza (2001) to 
question the HeI detection. In addition to the obvious problems caused by the 
crowding of features in the spectrum, the metallic shell lines have absorption 
and emission components, and the wavelength of the emission 
varies with orbital phase in a manner that is consistent with the motion of the 
accreting star (e.g. Figure 3). The emission from these metals 
might then be misinterpreted as originating from another element 
with a similar wavelength, such as HeI.

	Mean phase-binned spectra at wavelengths near 
HeI 6678 are shown in the left hand column of Figure 12, and the median of 
these is shown at the bottom of the panel. Shell lines bracket the expected 
location of HeI 6678. The right hand portion of Figure 12 shows the results 
of subtracting the median spectrum from each phase-binned spectrum, and 
there are subtle changes at wavelengths near the two 
shell lines with orbital phase that are likely due to the orbital 
motion of the donor star (Section 4).

\begin{figure}
\figurenum{12}
\plotone{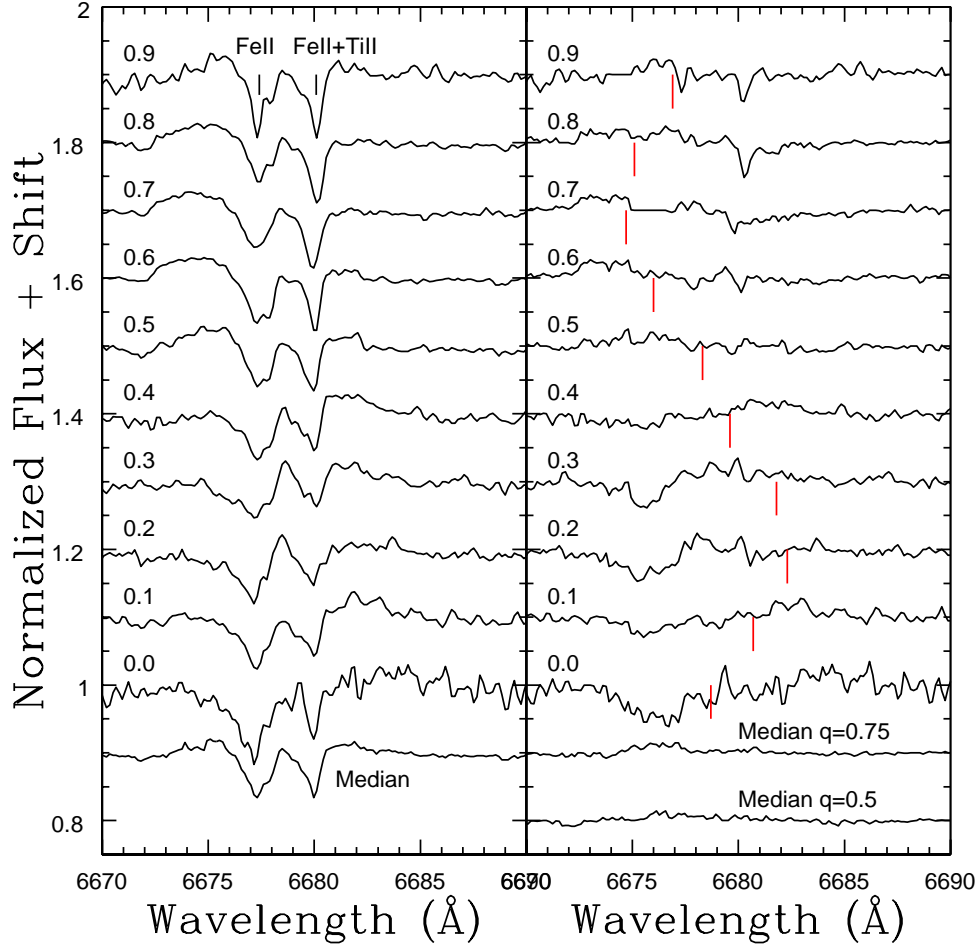}
\caption{Left hand panel: Spectra centered on HeI 6678. The median of 
spectra in all ten phase bins is shown at the bottom of the panel. 
Shell lines of FeII and TiII dominate at these wavelengths. 
Right hand panel: Residuals after subtracting the median spectrum from 
each spectrum in the left hand colmun. The red lines mark the 
expected location of HeI 6678 based on the velocity measurements made by 
Tarasov \& Berdyugin. Emission at a wavelength close to that of HeI 6678 in 
the restframe is seen in the residual spectra at phase 0.0. This 
feature is also clearly seen in the phase 0.0 
spectrum in the left hand panel. The median residual spectra 
after shifting spectra into the rest frame 
assuming mass ratios q=0.5 (i.e. that calculated by Tarasov 
\& Berdyugin) and 0.75 are shown at the bottom of the panel. The 
featureless nature of the combined spectra indicates that HeI 6678 is not 
detected throughout the entire orbital cycle.}
\end{figure}

	The vertical red lines in the right hand panel of Figure 12 mark the 
expected location of HeI at each phase if it followed the 
velocity curve measured by Tarasov \& Berdyugin (1998).
An emission line near 6679\AA\ is seen in the residual 
spectra at phase 0.0 (i.e. primary minimum), and this falls within the 
wavelength range expected for HeI in this phase 
bin. This line is also visible in the left 
hand panel, indicating that it is not an artifact of the removal of the median 
spectrum, but a real detection. We thus consider this to be a tentative 
detection of HeI 6678 at an orbital phase when the accreting star and its 
accompanying disk eclipse the donor star. The confirmation of this 
detection could be examined further with additional spectra of V367 Cyg, 
although with the caveat that if the hot spot moves on the accretion disk 
then the HeI line will appear at another phase.

	While our spectra hint that HeI 6678 is visible near 
primary minimum, it is not detected over a full orbital cycle. 
To demonstrate this, median spectra were constructed by aligning and 
combining the residual spectra according to the Tarasov \& Berdyugin 
velocities (i.e. a mass ratio of 0.5) as well as for a mass ratio of 0.75, 
and the results are shown at the bottom of the left hand panel. The combined 
spectra are featureless, with no feature near the expected HeI wavelength. 
We conclude that while there is possible evidence for 
HeI 6678 near primary minium, it is not detected throughout 
the orbital cycle of the system. This is consistent with the HeI 6678 
emission being powered by a source such as a hot spot that 
is not visible at all orbital phases.

\section{IR EMISSION AROUND V367 CYGNI}

	Mass transfer in a CBS may not be a conservative 
process. The loss of mass and angular momentum from the 
system hinders efforts to deduce the properties of the progenitor 
stars in present day Algol systems (e.g. Nelson 
\& Eggleton 2001), and to predict the final evolution of their more massive 
brethren (e.g. Kruckow et al. 2018). Van Rensbergen et al. (2011) find 
that the inclusion of mass and angular momentum loss in their models improves 
the agreement with the observed period distributions of Algols.

	Deschamps et al. (2015) model the IR properties of Algols that are in 
the early stages of mass transfer. Material is ejected from the system 
via a wind that originates in a hot spot that forms on the accretion disk. 
The ejected material cools as it expands outwards, and 
grains form in their baseline model at a distance 
of 212 AU from the system. A cloud of moderately hot grains spread over a 
large volume results. Their models indicate that IR emission from the cloud
may be detected with moderate-sized space-based telescopes over 
angular scales of many arcsec if observed at a distance of 300 pc.

	Excess IR emission from V367 Cyg was 
first noted by Taranova (1997), and entries in the WISE 
(Wright et al. 2010) point source catalogue at the IRSA 
\footnote[1]{https://wise2.ipac.caltech.edu/docs/release/allsky/src\_cat}
confirm that V367 Cyg is very bright at wavelengths out 
to at least $22\mu$m (i.e. the central wavelength of the W4 filter). 
Deschamps et al. (2015) note that V367 Cyg is not a point source 
in WISE images, with the most statistically significant deviation from 
a point source occuring in the W2 filter ($\lambda_{cen} \sim 4.5\mu$m).

	We have examined the morphology of the V367 Cyg envelope 
using W2 images that were downloaded from the WISE all-sky survey 
\footnote[2]{https://irsa.ipac.caltech.edu/applications/wise/}. 
The angular resolution of stars in the W2 image is $\sim 9$ 
arcsec FWHM, which corresponds to a spatial 
resolution of $\sim 6000$ AU at the distance of V367 Cyg.
The W2 image of V367 Cyg is compared with images of two stars that 
are in the same WISE image tile in Figure 13, and 
it is clear that V367 Cyg is not a point source. 

\begin{figure}
\figurenum{13}
\epsscale{0.5}
\plotone{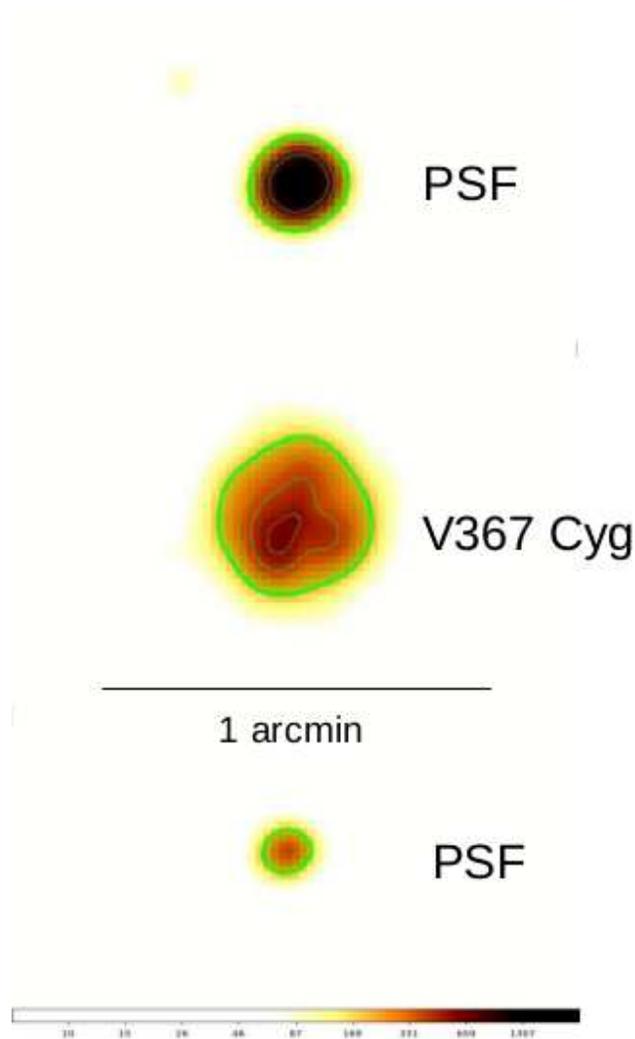}
\caption{W2 images of V367 Cyg and two stars in the same WISE image tile. 
All images are displayed with the same brightness stretch. The 
V367 Cyg isophotes are not flattened, suggesting that the emitting material 
is not confined to a disk. The FWHM of the stellar images is $\sim 9$ arcsec. 
V367 Cyg is not a point source, and there are localized peaks in the light 
distribution that may be companion stars. However, 
point sources alone can not account for the angular 
extent of the light distribution at $4.5\mu$m.}
\end{figure}

	The envelope around V367 Cyg is not flattened, even though the system 
is eclipsing, and so is viewed at an oblique angle. 
If the emission were restricted to a disk with an 
inclination of 80$^o$, as found from some light curve 
solutions (e.g. Zola \& Ogloza 2001), then the image would be flattened 
by $\sim 2:1$ after factoring in the angular resolution of the 
W2 images, which is clearly not the case. Taken at face value, 
the more-or-less round isophotes in Figure 13 suggest that the ejected 
material flares out over a larger volume than that defined by 
the orbital plane. 

	A caveat when investigating the morphology of the environment around 
V367 Cyg is that there is at least one moderately 
bright companion, and sub-structure is evident in 
the W2 image. Still, efforts to model the light 
distribution in Figure 13 as a collection of multiple point sources proved 
unsuccessful -- a diffuse background component centered on the system 
that has almost half of the signal level per pixel 
of the brightest source in Figure 13 is required to reproduce the 
extended nature of the W2 light distribution. Thus, while 
companions may be present, they can not explain the full extent of the 
light distribution at $4.5\mu$m. 

	More quantifiable insights into the spatial extent of 
emission at $4.5\mu$m can be gleaned from the light profile. 
Figure 14 shows the light profile through the middle column of V367 Cyg in 
Figure 13 along with the light profile of 
the PSF constructed from stellar images. 
Each profile has been normalized to its 
peak intensity. The V367 Cyg profile has a FWHM of 
almost 30 arcsec, as opposed to 9 arcsec for the PSF. After removing 
in quadrature the broadening produced by the PSF, then the 
intrinsic width of the W2 profile is 28 arcsec. This corresponds 
to 0.09 parsecs (19000 AU) if the GAIA-based distance of 670 
parsecs is adopted. Given the more-or-less symmetric shape of V367 Cyg in 
Figure 13, then selecting a different orientation for extracting a light 
profile should not yield a markedly different angular extent. Extended 
emission at $4.5\mu$m around V367 Cyg thus extends over spatial scales of 
at least a few percent of a parsec.

\begin{figure}
\figurenum{14}
\plotone{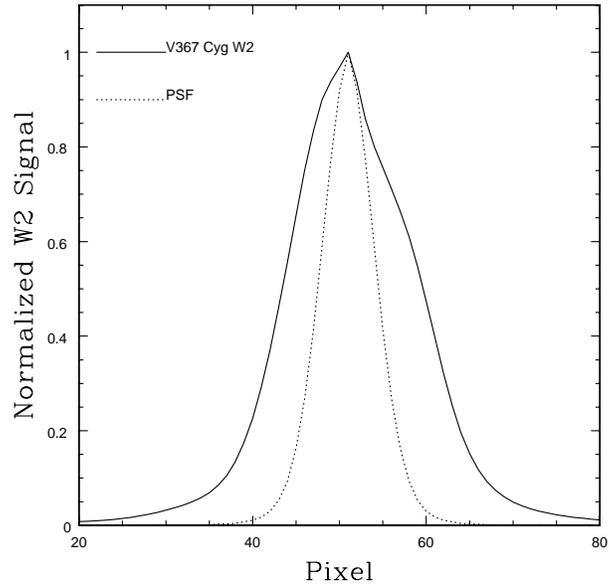}
\caption{Light profiles of V367 Cyg (solid line) and the PSF (dotted line) 
in the W2 filter. The profiles have been normalized to their peak values, and 
each pixel along the x axis subtends 1.5 arcsec. 
The V367 Cyg profile is that of the central 
column that passes through the image in Figure 13. 
The V367 Cyg light profile is wider than that 
of the PSF, and this can not be explained by 
multiple point sources (see text). Rather, there appears to be emitting 
material in the W2 image that is spread out over $\sim 19000$ AU 
around V367 Cyg.} 
\end{figure}
 
\section{SUMMARY \& DISCUSSION}

	Spectra that cover the wavelength interval 6320 -- 6900\AA\ 
with a resolution $\frac{\lambda}{\Delta \lambda} \sim 17000$ 
have been used to examine the CBS V367 Cyg and its surroundings. 
This wavelength interval has not been extensively observed in past 
studies of this system. The spectra were recorded over a 
7 month time span in 2021, and thus cover roughly 11 orbital cycles. 

	The observational characteristics of V367 Cyg suggest that it is a W 
Serpentis system, and so is likely undergoing the initial transfer of 
mass from the more evolved (and currently more massive) star to its companion. 
Van Rensbergen et al. (2011) model the orbital periods and mass ratio 
distributions of CBSs that contain a donor star that is initially spectral 
type B (i.e. a few solar masses), which is consistent with some of the 
mass estimates for V367 Cyg. The rate of mass transfer in 
their models increases towards longer orbital periods, 
since the donor star must be more evolved as the 
period increases in order for mass transfer to be initiated. These models 
predict that systems with orbital periods in excess of 15 
days will likely experience Case B mass transfer (i.e. the 
donor star has evolved off of the main sequence). 
To the extent that these models hold for V367 Cyg then it is likely 
that it is undergoing Case B mass transfer, and this is 
consistent with the spectroscopic characteristics of the 
donor star (Section 5).

	The deep and narrow absorption features that dominate the 
red spectrum originate in a circumsystem shell. 
There are sub-structures in the shell lines that change with 
orbital phase and time, although in Section 4 it is shown 
that there is a dominant absorption component in 
the shell lines that was more-or-less stable with time during the 
2021 observing season. Given that the shell almost certainly rotates, 
then this stability with time suggests that the main body 
of the shell has uniform structure, in agreement with models 
that predict that circumstellar material around mass-losing binary systems 
should be well mixed (e.g. Bermudez-Bustamante et al. 2020).

	Evidence is also presented in Section 4 that diffuse 
absorption lines that originate in the donor star spectrum are 
embedded in the much stronger shell lines, and that these 
lines likely have affected efforts to examine the dynamics  
of the circumsystem shell, in the sense 
that velocity measurements made from the shell lines have 
been skewed by contamination from the donor star spectrum. 
Previous studies have found that velocities measured from 
the shell lines define a radial velocity curve that is consistent with 
motion in the same direction as the donor star, but with an amplitude that 
is considerably smaller. However, the radial velocity curve 
constructed from a cross-correlation analysis of shell lines 
that is sensitive to the deepest parts of the strongest shell lines has 
an amplitude that is smaller than that found in previous studies. Moreover, 
velocities measured from only the deepest parts of the SiII 6347 and 
FeII 6432 shell lines, which is the part of the 
lines where contamination from the donor spectrum 
should least affect the velocity measurements, have an even smaller 
amplitude. We emphasize that the shell is likely rotating, but that it 
does not contain structures that affect the locations of line centers.

	The spectrum of the donor star has been extracted by shifting spectra 
with the shell lines removed into the restframe as defined by 
the orbital motion of the donor star and then combining the results. 
The mean spectrum has SiII and FeII absorption lines between 6300 and 6500\AA\ 
that are slightly weaker in strength than those in the spectrum of the A-type 
supergiant Deneb. This is consistent with V367 Cyg undergoing Case B mass 
transfer. The widths of the SiII 6347 line in the donor star spectrum indicate 
a rotation velocity of vsini $\sim 54$ km/sec, in agreement with earlier 
measurements made from MgII 4481. This rotation velocity is consistent with 
the donor star being tidally locked for some of the system geometries found 
by Zola \& Ogloza (2001).

	There is also evidence that the effective temperature 
across the face of the donor star varies with orbital phase, 
in the sense that SiII and FeII lines at phases 0.25 and 0.75 have 
different depths. Evidence for similar variations in the depth of metallic 
lines has been seen in previous studies at shorter wavelengths. 
That this difference in line strength occurs between phases 0.25 
and 0.75 is not consistent with gravity darkening, leading 
us to suspect that there is large scale spot activity on the donor star.  
That this behaviour has been evident over $\sim 50+$ years further suggests 
that the temperature difference between hemispheres 
has been in place for at least 1000 orbital cycles. 
This is consistent with the donor star being tidally locked, as suggested 
by its rotational velocity and some system geometries. Future 
observations of features such as Ca H and K 
or HeI 1.058$\mu$m would be useful to examine 
the nature of any spot activity on the donor.

	Given the success in extracting the spectrum of the donor star, 
an effort was also made to extract a spectrum associated with the 
accreting star. Shell-subtracted spectra were shifted into the rest frame using 
velocities that are appropriate for the accretor after assuming various system 
mass ratios. The results for each mass ratio were combined, 
and spectra constructed for mass ratios of unity 
and 0.75 contain emission features that are attributed to 
an accretion disk. These lines are formed from the emission components that are 
seen in the shoulders of metallic absorption lines. We caution that the 
emission lines in the resulting spectrum are weaker 
than their actual values, as the technique used to 
remove shell absorption lines suppresses signal from the emission line centers.

	The emission spectra constructed in this manner 
are similar in some respects to those seen in Herbig AeBe stars, 
and these objects have the potential to serve as an 
interpretive guide for understanding the emission spectrum of 
V367 Cyg. We emphasize that these observational similarities do not mean that 
we interpret the stars in V367 Cyg as young, forming objects. Rather, 
it is a pragmatic recognition that the physics of accretion activity in CBSs 
may produce observational similarities to other astrophysical objects in 
which accretions disks surround early-type stars -- in this case 
Herbig AeBe stars.

	It is perhaps not surprising that H$\alpha$ is 
by far the strongest feature in our spectra, appearing as 
a broad emission feature with deep central absorption. 
H$\alpha$ emission in CBSs is associated with interactions between 
the components, and simulations suggest that it should grow progressively 
in strength from Case A to Case C mass exchange as the 
mass flow increases (Deschamps et al. 2015). 
Previous studies have suggested that the donor 
star spectrum contributes detectable signatures in higher order Balmer lines. 
However, comparisons with the spectrum of Deneb 
indicate that only a very small fraction of the H$\alpha$ absorption 
can be attributed to the donor. Unlike was the case for shell lines, the 
donor star spectrum thus does not skew the H$\alpha$ profile. 

	The structure and strength of H$\alpha$ in 
the V367 Cyg spectrum changes with both time and orbital 
phase. The variations in strength with phase are in 
the sense that absorption and emission are most pronounced 
during the eclipses, suggesting a possible connection with the L2 and L3 
points. The deep central absorption makes it difficult 
to estimate radial velocities or velocity dispersions for the emission 
component, although the residuals in the right hand column of Figure 9 at 
phases outside of eclipse suggest that the emission moves in a 
manner that is consistent with the motion of the donor star. 
As for the absorption component, we speculate that 
structure in H$\alpha$ absorption might be due to discrete 
structures along the line of sight, such as might be expected if there is a
tightly wrapped spiral outflow. Modelling of such an outflow 
would be of interest to see if it can replicate the behaviour of H$\alpha$ 
absorption in these spectra.

	[OI] 6364 emission has been detected at all orbital phases. 
The detection of [OI] emission is perhaps not surprising 
given its presence in other accretion environments, such as Herbig AeBe stars. 
[OI] emission in Herbig AeBe stars is correlated with UV flux (Acke 
et al. 2005), which is a defining characteristic of W Ser systems.
[OI] emission also tends to be seen in Herbig AeBe stars that 
have an IR excess related to circumstellar emission (Viera et al. 2003; 
Acke et al. 2005), and excess IR emission is another common characteristic of 
W Ser stars. It would be of interest to survey the spectra of other W Ser 
systems to examine if there is a correlation between [OI] emission and UV 
flux and/or the presence of an IR excess.

	The [OI] 6364 feature is structurally complex, and 
has a shape that changes with orbital phase. 
The shape of the [OI] 6364 emission feature 
leads us to suspect that a large fraction of the flux 
is associated with the accretion disk, paralleling the interpretation of [OI] 
emission in the spectra of Herbig AeBe stars (e.g. Acke et al. 2005). 
[OI] 6364 should be accompanied by the much stronger [OI] 6300 line, which 
falls at the blue end of our wavelength coverage. Spectra that sample 6300\AA\ 
would be of interest to further characterize [OI] emission and probe 
conditions in the emitting region.

	The spectra have also been used to search for HeI 6678 emission, 
which might be expected given that it is seen in the model spectra 
of binary systems undergoing rapid mass transfer
generated by Deschamps et al. (2015). This being said, 
we do not find evidence of HeI emission over a wide range 
of orbital phases, contrary to what was found by Tarasov \& Berdyugin (1998). 
Rather, there is evidence for HeI emission only 
near primary minimum, suggesting that the emitting 
region is located on the side of the accreting star/disk that 
points away from the donor star. We caution that this detection is 
preliminary only given that the phase 0.0 spectrum is based on only one 
observation. We further note that HeI emission line 
might occur at different orbital phases in the future 
if the orientation of the excitation source in the system changes.

	What conclusions can be drawn about the future of V367 Cyg? 
An obvious problem is that the masses of the components 
are not firmly established. Still, while masses 
are at present uncertain, it might be possible to obtain an 
upper limit to the mass of the donor based on the environment around the 
system. Some of the mass estimates for the donor indicate that V367 Cyg 
may have formed within the past $\sim 10 - 20$ Myr. Given the 
rate at which star clusters are disrupted (e.g. Fall and Chandar 2012) 
then remnants of a star cluster might be found around the system. If 
such a remnant component could be identified through - say -- proper motion 
and/or parallax measurements then it might be possible to estimate an 
age for the cluster, and hence a plausible upper mass limit for the donor.

	Another source of uncertainty when predicting the future of 
V367 Cyg is that the system is shedding mass and angular 
momentum. The material is being fed into the surrounding 
ISM, with a dust envelope extending over at least 0.09 parsec. 
The non-conservative nature of the mass transfer complicates efforts to 
predict the future characteristics of the system given the uncertainties 
introduced by mass and angular momentum loss (e.g. Yakut \& Eggleton 2005).
These uncertainties notwithstanding, we suspect that V367 Cyg will evolve 
into a conventional Algol system after mass 
reversal, and the ensuing drop in the mass transfer rate. 
The mass transfer rate estimated by Zola \& Ogloza (2001) suggests that the 
mass ratio might reverse on time scales of $< 10^5$ years if the lower 
mass estimates for the component stars in Table 1 
are correct. The dust envelope is expected to dissipate once 
rapid mass transfer ends (Deschamps et al. 2015).

	We close by noting that the observations discussed here 
were recorded over only a few orbital cycles, and 
so provide a snap shot of the system in 2021. 
The evidence for large scale mass transfer and a circumsystem shell suggest 
that V367 Cyg is almost certainly in a rapid phase of its evolution, and so 
large-scale, transient changes in the observational 
characteristics of the system might occur over short timescales. 
Hints that this is happening are seen in the 
H$\alpha$ flux variations discussed in Section 5,
as well as the photometric variations in the light curve.

	Given the dynamic nature of the observational properties at the 
present day, longer term monitoring of V367 Cyg will undoubtedly prove 
to be rewarding, as it will provide a broader perspective of the 
nature of this system during a defining stage of 
its evolution. Obtaining spectra with a higher 
spectral resolution would be of particular interest to better 
decouple the various components that have been identified in this and 
earlier papers. This might in turn allow material around 
the accreting star to be detected throughout the full orbital cycle, 
raising the possibility of finding reliable masses for both components.

\acknowledgements{It is a pleasure to thank David Bohlender for reducing 
the spectra used in this paper. Thanks are also extended to Dmitry Monin 
for his tireless efforts configuring the DAO telescopes for these 
(and other) observing programs. This research 
has made use of the NASA/IPAC Infrared Science
Archive (https://doi.org/10.26131/irsa22), 
which is funded by the National Aeronautics and Space Administration and
operated by the California Institute of Technology.}

\appendix
\section{Shell Velocity Measurements}

	The shell velocities used to construct Figure 5 
are listed in Table A1. The estimated 
uncertainty in these measurements is $\pm 1$ km/sec.

\begin{table*}[ht]
\begin{center}
\begin{tabular}{crrrr}
\tableline\tableline
Phase & X-Corr\tablenotemark{a} & SiII 6347\tablenotemark{a} & FeII 6432\tablenotemark{a} & H$\alpha$\tablenotemark{b} \\
\tableline
0.0 & --3.3 & --10.3 & --5.1 & --3.7 \\
0.1 & --4.5 & --7.0 & --7.0 & --2.3 \\
0.2 & --5.7 & --9.3 & --7.9 & --3.2 \\
0.3 & --10.0 & --8.4 & --5.6 & 0.9 \\
0.4 & --8.3 & --6.5 & --6.5 & --1.4 \\
0.5 & 0.8 & --5.5 & --6.1 & --5.5 \\
0.6 & 6.4 & --1.3 & --2.8 & --3.7 \\
0.7 & 3.8 & 0.1 & --5.1 & --10.1 \\
0.8 & 7.3 & --0.3 & 0.5 & --3.2 \\
0.9 & 6.5 & --0.8 & --2.3 & --2.3 \\
\tableline
\end{tabular}
\end{center}
\caption{Shell Line and H$\alpha$ Velocity Measurements}
\tablenotetext{a}{Velocities of shell lines, in km/sec, measured from 
phase-binned spectra using the procedures discussed in Section 4.}
\tablenotetext{b}{Velocities of H$\alpha$ absorption, in km/sec, measured from 
phase-binned spectra using the procedures discussed in Section 6.}
\end{table*}

\end{document}